\documentclass[3p]{elsarticle}

\usepackage[utf8]{inputenc}
\usepackage[USenglish]{babel} 
\usepackage[T1]{fontenc}
\usepackage{lmodern} 
\usepackage{color}
\usepackage{mathtools}
\usepackage{graphicx} 

\usepackage{setspace}

\usepackage{amsmath}
\usepackage{amsthm}
\usepackage{amsfonts}

\newcommand{\bx}{{\bf x}}

\newcommand{\qh}[1]{q\!  \left(#1 \right) }
\newcommand{\q}[2]{q^{\phantom{*}}_{#1}\!  \left(#2 \right) }
\newcommand{\qc}[2]{q^*_{#1}\!  \left(#2 \right) }
\newcommand{\shat}{\sigma}

\newcommand{\pifac}{}
\newcommand{\intfac}[1]{{\frac{#1}{2\pi}}}
\newcommand{\iintfac}[1]{{\frac{#1}{(2 \pi)^2}}}

\newcommand{\Om}{{\omega}}

\newcommand{\knu}{{k(\nu)}}
\newcommand{\cnu}{{c(\nu)}}

\newcommand{\Omt}{{\tilde{\Om}}}
\newcommand{\Omp}{{\Om'}}

\newcommand{\Ha}[2]{\mathrm{H}_{#1}\!  \left(#2 \right) }
\newcommand{\I}{\mathrm{i}}
\newcommand{\E}{\mathrm{e}}

\newcommand{\Loeve}{Lo\`eve }

\newcommand{\dint}{\int\limits_{-\infty}^{\infty} \! \int\limits_{-\infty}^{\infty}}
\newcommand{\sint}{\int\limits_{-\infty}^{\infty} }

\newcommand{\bint}{\int\limits_{-\Om_B}^{\Om_B} }

\newcommand{\nuk}{\nu^{(k)}}
\newcommand{\nukcc}{\nu^{(k)*}}
\newcommand{\nukp}{\nu^{(k')}}
\newcommand{\nukpcc}{\nu^{(k')*}}

\newcommand{\qrint}{\int\limits_{\mathbb{R}^4} \! }

\newcommand{\vunit}{{m\,s$^{-1}$}}

\newcommand{\EE}[1]{\mathbb{E}\left[#1\right]}

\newcommand{\Plim}[1]{\underset{#1}{\text{P-lim}}}
\newcommand{\WO}{2 \pi W_f}
\newcommand{\wint}{\int\limits_{\mathcal{W}} }

\journal{J. Sound Vib.}
\begin{document}
\begin{frontmatter}
\title{Localizing broadband noise sources using the Lo\`eve spectrum and a 2.5D approach}
\author{Christian H. Kasess\fnref{ARI}}
\ead{christian.kasess@oeaw.ac.at}
\author{Wolfgang Kreuzer\fnref{ARI}}
\author{Holger Waubke\fnref{ARI}}

 \affiliation[ARI]{organization={Acoustics Research Institute, Austrian Academy of Sciences},
             addressline={Dominikanerbastei 16},
             city={Vienna},
             postcode={A-1010},
             country={Austria}}

\cortext[cor1]{Corresponding author: Christian H. Kasess}


\begin{abstract}
The localization of moving sound sources using a microphone array is typically based on modifying the signal to compensate for the Doppler effect.  In the time domain this compensation is done on a sample-by-sample basis. In the frequency domain short time segments need to be used in which the Doppler effect is assumed to be approximately constant and a discrete Fourier transform is done on each segment. 
In contrast, the authors developed  an inverse 2.5D localization method for uniformly moving single-frequency sources that works in the spectral domain and allows for the use of longer windows. This was achieved by modifying the 2.5D forward model to directly compute the effect of the motion in the static observer position. The method does neither require to modify the measured signal nor does it require quasi-stationary of the measurements within the window used. 
Unfortunately, this approach is not directly suitable for broad-band stochastic sources, and in the present work we will investigate how the statistical properties of a uniformly moving stochastic source change when observed at a static observer. 
Using a 2.5D setting, the relation between the power spectral density of the moving source and the Loève spectrum, which is a generalization of the cross-spectral density at the static receivers, was derived.  Based on simulated data with speeds up to 100 m\,s$^{-1}$, the work presented here provides a proof of concept for a method based on multi-taper estimates for the Loève spectrum to localize moving broad-band stochastic sources . Currently, the method requires a stationary source signal and that the spectral density is flat within a certain range around the frequency of interest. Also, correlations between sources are currently not considered. 
\end{abstract}

\end{frontmatter}
\pagestyle{empty} 

\pagestyle{plain} 

\section{Introduction}
Transportation noise is an important topic in acoustic research and the use of microphone arrays to  localize acoustic noise sources, which are often moving, has a long history of research. In particular, acoustic beamforming has received much attention and many different approaches have been and are still developed (see \cite{Merino2019,Chiariotti2019} for extensive reviews on the topic). 
In conventional beamforming approaches a set of so-called steering vectors is used to scan through a potential source space independently. Often, a second step is taken in which through a deconvolution the blurred beamforming map is "cleaned" to obtain a more focused source map \cite{Dougherty2005,Fleury2011,Sijtsma2007}. In contrast, inverse methods solve a joint optimization problem where this deconvolution step is an integral part of the solution process enforcing, e.g., sparseness via appropriate constraints (see, e.g., \cite{Suzuki2011,Meng2019}). Furthermore, inverse methods are better suited in situations where an  acoustic scatterers needs to be accounted \cite{Gombots2021,Schuhmacher2003}. A brief discussion on the difference between  beamforming and inverse approaches is given in the introduction of \cite{Kasess2024}.

For moving sources, localization methods mostly operate either purely in time-domain (see \cite{Cousson2019,Meng2019}) or in a hybrid setting. The latter use the time-domain to transform the problem from the static microphone array into the moving frame applying a de-dopplerization based on assumed source positions \cite{Guerin2008,Schumacher2022,Zhang2019,Zhang2023}. Afterwards, methods suitable for stationary signals, i.e. signals with time-independent properties, are used such as estimating the cross-spectral matrix.
Less commonly, pure frequency-domain approaches are used.  One example is the work on beamforming on linearly moving sources by Fleury et al. \cite{Fleury2011}, where short time windows were used to limit the source displacement and thus ensure a sufficient degree of stationarity in order to allow for an approximate Doppler compensation. For other types of motion, e.g., rotations see \cite{Chu2021,Gombots2021}.
 
In the present work we introduced a purely frequency based method for uniformly moving sources, i.e. sources moving along a straight line at constant speed. In previous work by the authors, an inverse localization approach for uniformly moving single-frequency sources based on the 2.5D Helmholtz boundary element method (BEM) was presented where the use of short time windows is not necessary anymore \cite{Kasess2024,Kasess2024b}. Under the assumption of a constant cross-section in the $y$-$z$ plane, the 2.5D approach allows to transform a problem in the 3D space into many 2D problems with different wave numbers. In contrast to pure 2D methods, this approach allows modeling of 3D point sources, incoherent line sources, or sources of finite length via a spatial Fourier transform along $x$. Once solutions of the different 2D problems have been derived, the 3D sound field is reconstructed using an inverse Fourier transform (see for example \cite{Duhamel1996,Fakhraei2022,Hornikx2007,Kamrath2018,Kasess2016,Pizarro-Ruiz2019,Wei2021,Velazquez-Mata2026}). Being a frequency-based method, the 2.5D Helmholtz BEM also allows for a straight-forward definition of impedance boundary conditions \cite{Duhamel1998}. Also, the 2.5D approach was chosen because it is particularly well suited to model sources moving along $x$ at a constant speed  (for a derivation see, e.g., \cite{Duhamel1996}).

In this previously presented method \cite{Kasess2024}, a single spectrum of the pass-by of a moving source was used for localization. However, the model was restricted to single-frequency sources to reduce the number of calculations. The leakage effect of the discrete Fourier transform (DFT) could be accommodated in the forward model and lead to an improvement in the localization performance, in particular for shorter observation periods. 
This 2.5D approach was also shown to be applicable in a more general setting modeling a scattering structure as well as the source structure itself \cite{Kasess2024b}.
A case in point for such a setting is a microphone array measurement of a moving high-speed train where the array is positioned close to the track leading to highly transient signals. Typical noise sources in this scenario are, e.g., aerodynamic sources and rolling noise. The latter can be seen as a stochastic process generated by combined wheel and rail roughness which is filtered via different transfer paths \cite{Thompson2009}. Thus, to overcome the restriction to single-frequency sources, the main novelty of the work presented here is an extension of \cite{Kasess2024} to stochastic broad-band sources.

The new idea presented here is to use the 2.5D framework for uniformly moving sources to derive the transformation of certain statistical quantities of the moving stochastic source process to the static receiver array. Thus, instead of trying to obtain stationary signal using some transformation in the time domain, the \Loeve spectrum will be used which is a generalization of the (cross)-spectral density. In contrast to the latter, the \Loeve is defined in two frequencies and is a 2D Fourier transform of the (cross)-covariance function for non-stationary processes. 

To motivate the use of the \Loeve spectrum, first a short outline of the main theoretical principles for harmonizable stochastic processes \cite{Loeve1963,Cramer1951}, which are a generalization of stationary random processes, will be provided in Sec.~\ref{Sec:Harm}. This type of process is required as the observed signals at the static receiver array are clearly not stationary as their properties change over time due to the Doppler effect. Furthermore, to obtain the \Loeve spectrum for actual signals, an existing estimator based on the multitaper method will be used and described in detail \cite{Thomson1982,Thomson2001}.
Note that to avoid confusions, the term stationary will only be used in the statistical sense and will always imply time-independent statistical properties whereas the term static will be used as the opposite of moving.

In Sec.~\ref{Sec:moving} the theoretical relation between the source and the observed process for a general 2.5D setting which may include scatterers and stochastic broad band source signals will be derived to show that the \Loeve spectrum can be directly related to the 2.5D framework. To reduce the computational effort a number of assumptions on the source process will be motivated. First, a wide-sense stationary (WSS) source process is assumed. 
Furthermore, a flat spectrum over a certain frequency range and uncorrelated sources are assumed. Similar to the consideration of the leakage effect in \cite{Kasess2024} it will also be shown how the estimator used affects the theoretical \Loeve spectrum. 

Note that the current work focuses on the forward problem, i.e. the theoretical transformation of the \Loeve spectrum and the relation to a particular estimator. Suitable methods for the inversion of the system will not be considered here. Based on simulated data of a moving source in a free field propagation a proof of concept will be provided to show that the presented approach allows to localize and distinguish different moving sources. The potential effects of having a non-flat spectrum and using correlated sources will be illustrated and discussed.

\section{Harmonizable process and the \Loeve spectrum}\label{Sec:Harm}
Before deriving a 2.5D method for uniformly moving stochastic sound sources (i.e., sources with stochastic frequency spectrum) a brief overview over harmonizable, stationary, and non-stationary random processes will be given in this section. As the use of radial frequencies $\Om=2 \pi f$ and wavenumbers $k=\Om c^{-1}$ ($c$ denotes the speed of sound in the medium) is more common in the 2.5D setting, the Fourier integrals necessary for the method will be defined in terms of these quantities. Continuous random variables in the time domain will be denoted by capital $X$, whereas $Y$ or $Z$ will be used for the frequency representation of such a process. Multivariate random processes are indicated by indices $\ell$ and/or $n$. Random sequences will be denoted similarly as, e.g., $X[t_j]$ where the brackets indicate the discrete nature of the argument. The time samples are defined as $t_j=j f_s^{-1}$ with $f_s$ being the sampling frequency. For realizations of random processes the respective lower case letters will be used. Note that this leads to an ambiguity regarding the spatial coordinate system. However, the distinction is clear from the context. The imaginary unit is defined as $\I=+\sqrt{-1}$.
Expectation values will be denoted as $\EE{\cdot}$ and estimators will be  indicated using a hat '$\hat{\cdot}$' symbol.

\subsection{Harmonizable random process}
If a time continuous random process $X(t)$ is harmonizable then it has the following integral representation 
\begin{align}\label{Equ:RProc}
X(t) &= \intfac{1} \sint \E^{-\I \Om t} dZ(\Om),
\end{align}
where $dZ(\Om)$ is a random measure and is sometimes also called the generalized Fourier transform of the process \cite{Thomson2001}.\footnote{Note that in this manuscript we follow the sign convention for the Fourier transform used in \cite{Duhamel1996}.} 

In \cite{Loeve1963,Cramer1951,Cambanis1970} it was shown that the process $X(t)$ is harmonizable if and only if the following relation for the autocovariance function holds:
\begin{align}\label{Equ:RProcCov}
\nonumber r(t,t') &= \EE{X(t) X^*(t')}\} \\
\nonumber &= \iintfac{1} \dint \E^{-\I \Om t + \I \Om' t'} \EE{dZ(\Om) dZ^*(\Om')}\\
&= \iintfac{1} \dint \E^{-\I \Om t + \I \Om' t'} \gamma(\Om,\Om') d\Om d\Om',
\end{align}
where the generalized spectral density is given as $\EE{dZ(\Om) dZ^*(\Om')}=\gamma(\Om,\Om') d\Om d\Om'$ and  $\gamma(\Om,\Om')$ is called the \Loeve spectrum \cite{Thomson2001}. The star denotes the complex conjugate.
For a wide sense stationary (WSS) process, $r(t,t')$ is a function of $t-t'$ only which is equivalent to a \Loeve spectrum of the following form:
\begin{align}\label{Equ:RProcCov1}
\gamma(\Om,\Om') &= 2\pi S_X(\Om) \delta(\Om-\Om').
\end{align}
$S_X$ is the power spectral density (PSD) of the process $X(t)$ and $\delta(\cdot)$ is the delta distribution or Dirac delta. 
Loosely speaking, for a stationary process different frequency bands are uncorrelated in its spectral representation.

Similarly, a harmonizable \emph{discrete-time} process can be defined as
\begin{align}\label{Equ:RProcDiscrete}
X[t_j] &= \intfac{1} \bint \E^{-\I \Om t_j} dZ_B(\Om),
\end{align}
where $dZ_B(\Om)$ has to be band limited to $\pm \Om_B$ and $\Om_B<\pi f_s$ to fulfill the Shannon sampling theorem in order to guarantee proper sampling accuracy. However, note that the processes under consideration here are not necessarily strictly band limited but they will be assumed to be decaying sufficiently fast not to exhibit any substantial contribution beyond $\pm \Om_B$. 

For the remainder of the manuscript all processes will be considered to be band limited even for the time-continuous case. To avoid a cluttered notation, the subscript $B$ indicating the band limited case will be dropped. 
\subsection{Important related quantities}
A common transformation is to rotate the frequency axes by 45°, i.e., substituting $\Om_s=0.5 (\Om+\Om')$ and $\Om_n=\Om-\Om'$. The rotated spectrum is then defined as $\tilde{\gamma}(\Om_n,\Om_s) = \gamma(\Om_s-\Om_n/2,\Om_s+\Om_n/2)$. 
These transformed frequencies are also sometimes termed the \textit{ordinary} or \textit{stationary} frequency $\Om_s$ and the \textit{non-stationary} frequency $\Om_n$. 
Similarly, defining a time $t_0=0.5(t+t')$ and a lag $\tau = t-t'$ the correlation function in Eq.~(\ref{Equ:RProcCov}) can be redefined as 
\begin{align}\label{Equ:RProcCovRot}
\nonumber \tilde{r}(t_0,\tau) &= r(t_0+\tau/2,t_0-\tau/2) \\
\nonumber		&= \iintfac{1} \dint \E^{-\I \Om (t_0+\tau/2) + \I \Om' (t_0-\tau/2)} \gamma(\Om,\Om') d\Om d\Om'\\
				&= \iintfac{1} \dint \E^{-\I \Om_s \tau - \I \Om_n t_0} \tilde{\gamma}(\Om_n,\Om_s) d\Om_s d\Om_n,
\end{align}
where in the last step a change of variables from $(\Om,\Om')$ to $(\Om_s,\Om_t)$ was performed. Loosely speaking, the ordinary frequency $\Om_s$ is associated with the time lag $\tau$ whereas $\Om_n$ is related to the absolute time $t_0$. A stationary process in this representation would be confined to the  $\Om_s$-axis, hence the name 'stationary frequency':
\begin{align}\label{Equ:RProcCov1Rot}
\tilde{\gamma}(\Om_s,\Om_n) &= 2 \pi S_X(\Om_s) \delta(\Om_n).
\end{align}
This expression is equivalent to Eq.~(\ref{Equ:RProcCov1}). This transformation will be important for understanding the properties of the estimator in Sec.~\ref{Sec:Est}. It is also strongly related to other quantities describing harmonizable processes.
Taking the Fourier transform of $\tilde{r}(t_0,\tau)$ w.r.t. $\tau$ or the inverse Fourier transform of $\tilde{\gamma}(\Om_n,\Om_s)$ w.r.t. $\Om_n$ leads to the so called Wigner-Ville-Spectrum of $X(t)$, which in turn is related to the spectrogram of the process (see \cite{Bayram2001}). In contrast, taking the Fourier transform of $\tilde{r}(t_0,\tau)$ w.r.t. $t_0$ or the inverse Fourier transform of $\tilde{\gamma}(\Om_n,\Om_s)$ w.r.t. $\Om_s$ leads to the ambiguity function, a quantity particularly important for radar (see, e.g., \cite{Zhou2023}).

The choice of using the \Loeve spectrum instead of other quantities will be motivated in Sec.~\ref{Sec:moving}, where the transformation of a harmonizable process in the 2.5D framework will be derived.

\subsection{Multitaper Spectral Estimator}\label{Sec:Est}
For stationary signals the cross-spectral density is typically used in spectral beamforming and this density is only truly meaningful for stationary processes. Most commonly, a Welch-type estimator is  used, which divides the data into short consecutive segments with a temporal overlap. For each segment spectral density estimates are gained using a windowed Discrete Fourier Transform (wDFT), i.e., the segment is multiplied with a window function often also referred to as a taper before the DFT typically in order to achieve better spectral properties such as reduced leakage. For stationary processes the spectral density estimates are simply averaged over time. 

An alternative approach that will be used here was introduced by Thomson in  \cite{Thomson1982} where instead of dividing the data into short segments, a number of different orthogonal window functions or tapers are used on the full segment, hence the name multitaper method or MTM. The typical set of tapers for spectral estimation are the discrete prolate spheroidal sequences (DPSS or Slepian sequences) because of their optimal properties for spectral analysis. 
The main motivation for using the DPSS is the well known fact in signal theory that a band limited function cannot have finite support in time. However one can look for finite sequences that have optimal energy concentration in a given band $[-W,W]$, or vice versa, look for band limited functions that have their energy optimally concentrated in a given time interval (concentration problem). In \cite{Slepian1978} the prolate spheroidal wave functions were derived as a solution to that concentration problem. 

Briefly, the prolate spheroidal wave functions or Slepian functions are the eigenfunctions of the Dirichlet kernel for a predefined bandwidth $W$. They are related to the DPSS on the interval $[0,M-1]$ via the Discrete Time Fourier Transform (DTFT) and are maximally concentrated in the frequency band given by $W$. Since the DPSS are only used on the interval $[0,M-1]$, using the DTFT implies a zero padding outside this interval and is thus equivalent to a sinc interpolation for frequency bins not on the discrete frequency grid of the DFT with length $M$.
Importantly, the DPSS are orthonormal on $[0,M-1]$ as well as orthogonal in $\mathbb{Z}$. Thus, instead of using temporally divided data segments, a set of different DPSS on $[0,M-1]$ can be used and averaged across. Depending on the desired bandwidth $W$ and the length $M$, the number of tapers needed varies, as will be outlined briefly now. 
While the DPSS and the Slepian functions are 'optimal' for spectral analysis, for other scenarios other functions may be more suitable. For example, for time-frequency analysis, the Hermite functions were found to be the ideal representation \cite{Bayram2001}. 

For spectral analysis, starting with measurements of a time-discrete multivariate process of dimension $N$ the wDFT of the $n$-th component $x_n[t_m]$ is given as
\begin{align}
z_n^{(k)}[\Om] = \sum\limits_{m=0}^{M-1} v^{(k)}[t_m] \, x_n[t_m] \E^{\I \Om t_m },
\end{align}
where the window function or taper $v^{(k)}$ is the $k$-th DPSS of length $M=T f_s^{-1}$ samples where $T$ is the duration or length of the taper and $f_s$ is the sampling frequency.
Note that this is an abbreviated notation, as strictly speaking $v^{(k)}[t_m]$ is parameterized by the bandwidth $W$ and the number of samples $M$.
The brackets $[\cdot]$ indicate that the quantity is a sequence and with the time or frequency argument being from a discrete set with $t_m=m f_s^{-1}$  
and $\Om$ is an integer multiple of $2 \pi  f_s M^{-1}$.

Please note that in a slight deviation of notation, no index is used on $\Om$ to avoid introducing another set of index variables. Furthermore, in theory off-grid frequencies could also be calculated using zero-padding in the time domain for the DFT. However, here all examples will be restricted to the frequency grid given by $M$ and $f_s$.

In the stationary case the estimator for the PSD is given as
\begin{align}
\widehat{S}_n[\Om] = \frac{1}{K}\sum\limits_{k=0}^{K-1} \widehat{z}_n^{(k)}[\Om] \, \widehat{z}_n^{(k)*}[\Om].
\end{align}
Note, that instead of the raw DFT coefficients $z_n^{(k)}[\Om]$ estimated quantities $\widehat{z}_n^{(k)}[\Om]$ are used indicated by the hat symbol.  The simplest estimator is to use $\widehat{z}_n^{(k)}[\Om]=z_n^{(k)}[\Om]$ directly. 
However, various procedures exist to weight or adjust the coefficients in order to reduce bias caused by the spectral leakage in particular of the higher-order tapers (see, e.g., \cite[Sec.~3.3]{Thomson2001}) and thus, in general $\widehat{z}_n^{(k)}[\Om] \neq z_n^{(k)}[\Om]$. 

The number $K$ of used DPSSs is determined depending on the product $MW$ where $2W$ is the bandwidth of the Slepians in the normalized frequency domain. 
Typical values for the product $MW$ are 2, 4, or 8 and the number of DPSS used  is $K=2MW$ \cite[Eq.~(59)]{Slepian1978}. 
The latter ensures that the associated eigenvalues are close to one which means that the energy of the $k$-th window is mainly concentrated in the frequency band defined by $W$. Sometimes lower numbers of tapers, e.g., $2MW-2$ or $2MW-4$ are used to avoid the increased leakage of the higher order tapers in the transition zone of the eigenvalues. However, this may come at the expense of an increased variance \cite{Thomson2001}.
The half bandwidth $W_f$ in Hertz for a given $K$ and $M$ can be calculated as $W_f=W f_s = 0.5 K M^{-1} f_s = 0.5 K T^{-1}$, where $T$ is the length of the DPSS in seconds.

Next, the estimation of the \Loeve spectrum of measured data is treated.  In the original multitaper publication \cite{Thomson1982} an estimator for the \Loeve spectrum was introduced as
\begin{align}\label{Equ:LoeveStatMT} 
\widehat{\gamma}_{n n'}[\Om,\Om'] = \frac{1}{K}\sum\limits_{k=0}^{K-1} \widehat{z}_n^{(k)}[\Om] \, \widehat{z}_{n'}^{(k)*}[\Om'].
\end{align}
In previous work \cite{Kasess2024} it was shown that spectral leakage due to the wDFT could affect the localization performance in the single frequency case. 
For this, in a first step it is important to understand how the above estimator is derived. In the following a brief summary and a motivation for Eq.~(\ref{Equ:LoeveStatMT}) above is given that is taken from \cite{Thomson2001}. 

The observable portion of the spectral representation $dZ_n(\Om)$ of a process $X_n(t)$ within a certain band $\mathcal{W}=(-\WO,\WO)$ is defined using the following expansion \cite[Eq.~(3.20)]{Thomson2001}:
\begin{align}\label{Equ:Exp1}
d\tilde{Z}_n(\Om + \xi) &\approx \sum\limits_{k=0}^{K-1} Z_n^{(k)}[\Om] \, \nuk(\xi) d\xi.
\end{align}
Thomson states that observable means the recoverable portion of $dZ_n$, if it were known, using an $M$-point Slepian. 
$\nuk$ denotes the $k$-th Slepian wave functions, i.e. the DTFT of $v^{(k)}$, which is continuous in frequency.   
The coefficients $Z^{(k)}[\omega]$ are defined as in \cite[Eq.~(3.21)]{Thomson2001}:
\begin{align}\label{Equ:Exp2}
 Z_n^{(k)}[\Om]   &= \intfac{1} \wint \nukcc(\xi) \,dZ_n(\Om + \xi). 
\end{align}
The capital letter indicates, that these 'idealized' coefficients, which are unobservable, are still stochastic quantities. As all  $\nuk(\xi)$ are orthonormal on the interval $\mathcal{W}$, Eqs.~(\ref{Equ:Exp1}) and (\ref{Equ:Exp2}) can be interpreted as a projection of $dZ_n$ onto the subspace spanned by $\nuk(\xi), k = 0,\dots K-1$ around a fixed $\Om$. Since the $\nuk(\xi), k = 0,\dots K-1$ are assumed to have most of their energy concentrated in $\mathcal{W}$ this projection can be loosely interpreted as a projection onto $(\Om- \WO,\Om+\WO)$. 
Conversely, using the actual estimates leads to the expansion
\begin{align}\label{Equ:ExpEst1}
d\widehat{Z}_n(\Om +\xi) &= \sum\limits_{k=0}^{K-1} \widehat{z}_n^{(k)}[\Om] \, \nuk(\xi) d\xi.
\end{align}
The estimator in Eq.~(\ref{Equ:LoeveStatMT}) is based on this expansion and a weighting function $H(\xi,\xi')$ in the following way \cite[Sec. 7]{Thomson2001}:
\begin{align}\label{Equ:LoeveEst}
\nonumber \widehat{\gamma}_{n n'}[\Om,\Om']  &=  \frac{1}{K} (2 \pi)^{-2} \wint\wint H(\xi,\xi') d\widehat{Z}_{n}(\Om+\xi) d\widehat{Z}_{n'}^*(\Om'+\xi') \\ 
&= \frac{1}{K} (2 \pi)^{-2} \wint\wint H(\xi,\xi') \sum\limits_{k,k'=0}^{K-1} \widehat{z}^{(k)}[\Om]\, \widehat{z}^{(k')*}[\Omp] \, \nukcc(\xi) \nukp(\xi') d\xi d\xi',
\end{align}
where Eq.~(\ref{Equ:ExpEst1}) was used.

Besides $\hat{z}^{(k)}$ the estimator for the \Loeve spectrum is defined by the functional form of the weighting function $H$. The estimator in Eq.~(\ref{Equ:LoeveStatMT})  originates from a weighting function $H(\xi,\xi')=2 \pi \delta(\xi - \xi')$. Conceptionally, this implies smoothing over the stationary frequency and no smoothing over the non-stationary frequency axis defined in Eq.~(\ref{Equ:RProcCovRot}).
Using Eq.~(\ref{Equ:LoeveStatMT}), applying the delta distribution, and using the orthogonality relation
\begin{align}\label{Equ:Ortho}
 \wint \nukcc(\xi)  \nukp(\xi) d\xi&= 2 \pi \delta_{kk'} 
\end{align}
yields the estimator in Eq.~(\ref{Equ:LoeveStatMT}).
There are, of course, other estimators for the \Loeve spectrum based on different approaches. For example, the above mentioned multi-taper approach for the time-frequency plane \cite{Bayram2001} can be used to estimate the Wigner-Ville-Spectrum which in turn, is used to estimate the \Loeve spectrum via a DFT \cite{Huang2024}. For a brief list of other estimators for the \Loeve spectrum please refer to that same publication.

\section{2.5D approach for a moving stochastic source}\label{Sec:moving}
\subsection{Moving stochastic source}
The starting point for modeling moving stochastic sources will be the formulation presented in \cite[Eq.~(50)]{Duhamel1996}. Briefly, a single (point) source emitting a source signal $s(t)$ in the time domain is assumed which moves at a constant speed of $v_s$ along the $x$-axis: $x_s(t) = x_{0} + v_s t$. 

Using a source at $\bx_s = \left(x_s(t),y_s,z_s\right)^\top$, the pressure at the receiver position $\bx_r = \left(x_r,y_r,z_r\right)^\top$ at time $t$ is given as:
\begin{align}\label{Equ:Uni}
p(x_{0},y_s,z_s,x_r,y_r,z_r,t) &= \iintfac{1} \dint  \shat(\Om - v_s k_x) \qh{y_s,z_s,y_r,z_r,\sqrt{\Om^2/c^2-k^2_x}}  \E^{\I  k_x (x_r - x_{0})} \E^{-i \pifac \Om t} dk_x  d\Om.
\end{align}
$\shat(\Om - v_s k_x)$ is the Fourier transform of $s(t)$ evaluated at $\Om - v_s k_x$.

For a single moving pressure point source in the free field $\qh{y_s,z_s,y_r,z_r,k_2} = \frac{\I}{4} \Ha{0}{k_2 r_2 }$ and $r_2 := \sqrt{(y_r-y_s)^2+(z_r-z_s)^2}, k_2 := \sqrt{\Om^2/c^2 - k_x^2}$. In a more general setting, $\qh{\cdot}$ is a solution of the 2D Helmholtz equation for the 2D wave number $k_2$.

The most common situation in source localization is the use of a microphone array with $N$ microphones and a grid of $L$ potential sources / source positions. In an inverse BEM approach the source positions are estimated using a comparison between measured and calculated data at the different microphone positions. 
Under the assumptions of $N$ receivers and $L$ potential sources the 3D sound pressure $p$ and the 2D BEM solutions $q$ are matrix functions of dimension $N \times L$. 
Indexing a potential source using $\ell$ with $\ell\in\{0, \dots, L-1\}$ and a potential receiver using $n$ with $n\in\{0, \dots, N-1\}$ Eq.~(\ref{Equ:Uni}) becomes
\begin{align}\label{Equ:UniInd} 
 p_{n\ell}(t) &= \iintfac{1} \dint  \shat_\ell(\Om - v_s k_x) \q{n\ell}{\sqrt{\Om^2/c^2-k^2_x}}  \E^{\I k_x (x_{r,n} - x_{s,\ell})} \E^{-\I \Om t} dk_x  d\Om,
\end{align}
where $x_{s,\ell}$ is the $x$-position of the $\ell$-th source at $t=0$. $\q{n\ell\phantom{'}}{\cdot}$ is now, among other dependencies, a function of the $y$ and $z$-coordinates of the respective source (index $\ell$) and receiver (index $n$) position but not of the $x$-coordinates which are encoded in an oscillatory term.

For reasons of symmetry that will become clear later we define  $\Omt := \Om - v_s k_x$,  leading to
\begin{align}\label{Equ:UniInd2} 
 p_{n\ell}(t) &= \frac{1}{v_s (2 \pi)^2} \dint  \shat_\ell(\Omt) \q{n\ell}{\sqrt{\Om^2/c^2-(\Om-\Omt)^2/v_s^2}}  \E^{\I (\Om-\Omt) v_s^{-1} (x_{r,n} - x_{s,\ell})} \E^{-\I \Om t} d\Omt  d\Om.
\end{align}
Now, $s_{\ell}(t)$ is taken to be a harmonizable random process. Thereby, $\shat_\ell(\Omt) d\Omt$ is replaced by $dY_\ell(\Omt)$. For details on this step see Appendix A. Then, Eq.~(\ref{Equ:UniInd2}) reads as:
\begin{align}\label{Equ:Unirand} 
\nonumber X_{n\ell}(t) &= \frac{1}{v_s(2 \pi)^2} \dint  \q{n\ell}{\sqrt{\Om^2/c^2-(\Om-\Omt)^2/v_s^2}}  \E^{\I (\Om-\Omt) v_s^{-1} (x_{r,n} - x_{s,\ell})} \E^{-\I \Om t} dY_\ell(\Omt)  d\Om\\
&=\frac{1}{v_s(2 \pi)^2} \dint  \q{n\ell}{\Om,\Omt}  \E^{\I (\Om-\Omt) v_s^{-1} (x_{n} - x_{\ell})} \E^{-\I \Om t} dY_\ell(\Omt)  d\Om,
\end{align}
where a shorthand notation for $\q{n \ell}{\cdot}$ is introduced. To further shorten the notation, the indices '$r$' for receiver and '$s$' for source will be dropped in the following. 
Note that  $Y$ is used to indicate that this is not the generalized spectrum of the process $X_{n\ell}(t)$ observed at the static receiver but rather of the unknown source process. Strictly speaking, the notation after introducing the stochastic process is a simplification for which the reasons are outlined in Appendix A.

Using the above expression, the covariance function of the observed signals between the different measurement positions with $x$-coordinates $x_n$ and $x_{n'}$ induced by sources with $x$-positions $x_\ell$ and $x_{\ell'}$ as
\begin{align}\label{Equ:UniCorr2} 
&r_{nn'\ell\ell'}(t,t') = \nonumber \EE{X_{n\ell}(t) X_{n'\ell'}^*(t')} \\
 \nonumber &= \frac{1}{v_s^2(2  \pi)^4}\! \qrint \EE{dY_\ell(\Omt) dY^*_{\ell'}(\Omt') } \q{n \ell}{\Om,\Omt} \qc{n' \ell'}{\Om',\Omt'} 
 \E^{\I (\Om-\Omt) v_s^{-1} (x_{n} - x_{\ell})} \E^{-\I (\Om' - \Omt') v_s^{-1}(x_{n'} - x_{\ell'})} \E^{-\I \Om t} \E^{\I \Om' t'}  d\Om d\Om'\\
&= \frac{1}{v_s^2(2  \pi)^4}\! \qrint \gamma_{\ell \ell'}(\Omt,\Omt') \q{n \ell}{\Om,\Omt} \qc{n' \ell'}{\Om',\Omt'} 
\E^{\I (\Om-\Omt) v_s^{-1} (x_{n} - x_{\ell})} \E^{-\I (\Om'-\Omt') v_s^{-1} (x_{n'} - x_{\ell'})}\E^{-\I \Om t} \E^{\I \Om' t'} d\Omt d\Omt' d\Om d\Om'.
\end{align}
where the last step was the same as in Eq.~(\ref{Equ:RProcCov}). 
Clearly, to calculate this expression is numerically challenging.
Assuming that the observed multivariate process is also harmonizable, the \Loeve spectrum  $\gamma_{nn' \ell\ell'}(\Om,\Om')$ can be derived by the Fourier transform of Eq.~(\ref{Equ:UniCorr2}), i.e., using the relation given in Eq.~(\ref{Equ:RProcCov}): 
\begin{align}\label{Equ:UniLoeve1} 
\gamma_{nn' \ell\ell'}(\Om,\Om') =\frac{1}{v_s^2 (2  \pi)^2} \dint \q{n\ell}{\Om,\Omt} \qc{n' \ell'}{\Om',\Omt'}   \E^{\I (\Om-\Omt) v_s^{-1} (x_n - x_\ell)} \E^{-\I (\Om' - \Omt') v_s^{-1} (x_{n'} - x_{\ell'})} \gamma_{\ell \ell'}(\Omt,\Omt') d\Omt d\Omt'. 
\end{align}
At this point it is important to notice, that for practical situations including scattering structures each evaluation of $\qh{\cdot}$ involves a 2D BE calculation. Thus, the numerical calculation of Eq.~(\ref{Equ:UniLoeve1}) involves a large number of  2D BE calculations.  The next sections deal with possible simplifications under certain assumptions to reduce the number of said calculations.

\subsection{Wide-sense stationary source}
Up to now, no restrictions were placed on the source process. A significant simplification of Eq.~(\ref{Equ:UniLoeve1}) can be achieved by assuming that the source process is wide-sense stationary (WSS). 
For a multivariate WSS source process (see Eq.~(\ref{Equ:RProcCov1}) for the definition) the \Loeve spectrum at the observer becomes
\begin{align}\label{Equ:MRProcCov1} 
\gamma_{\ell \ell'}(\Omt,\Omt')= S_{\ell\ell'}(\Omt) 2 \pi \delta(\Omt-\Omt'),
\end{align}
where the statistical properties are determined by the cross-spectral density matrix $S_{\ell \ell'}(\Omt)$ of the (multivariate) stationary source process indexed by $\ell$ and $\ell'$. Using this expression in Eq.~(\ref{Equ:UniLoeve1}) leads to
\begin{align}
\nonumber &\gamma_{nn' \ell\ell'}(\Om,\Om')=\\
\nonumber  &= \frac{1}{v_s^2 2  \pi} \dint \q{n\ell}{\Om,\Omt} \qc{n' \ell'}{\Om',\Omt'}   \E^{\I (\Om-\Omt) v_s^{-1} (x_n - x_\ell)} \E^{-\I (\Om' - \Omt') v_s^{-1} (x_{n'} - x_{\ell'})} S_{\ell\ell'}(\Omt) \delta(\Omt-\Omt') d\Omt d\Omt' \\
\nonumber &= \frac{1}{v_s^2 2  \pi} \sint \q{n\ell}{\Om,\Omt} \qc{n' \ell'}{\Om',\Omt}  \E^{\I (\Om-\Omt) v_s^{-1} (x_n - x_\ell)} \E^{-\I (\Om' - \Omt) v_s^{-1} (x_{n'} - x_{\ell'})} S_{\ell\ell'}(\Omt) d\Omt \\
&= \frac{1}{v_s^2 2  \pi} \E^{\I \Om v_s^{-1} (x_n - x_\ell)} \E^{-\I \Om' v_s^{-1} (x_{n'} - x_{\ell'})}  \sint \q{n\ell}{\Om,\Omt} \qc{n' \ell'}{\Om',\Omt}  \E^{-\I\Omt v_s^{-1} (x_n - x_{n'} - x_\ell  + x_{\ell'})}  S_{\ell \ell'}(\Omt) d\Omt,
\end{align}
where in the last step all constant terms were brought in  front of the integral. Assuming stationary source processes allows to reduce the calculation of the \Loeve spectrum measured at receivers $n$ and $n'$ caused by sources $\ell$ and $\ell'$ to a one-dimensional integral. Clearly, due to the motion the process observed at the receiver array is not stationary anymore leading to a coupling between different frequencies.

\subsection{Cross-terms of the source}
When the dimension of the source grid $L$ is large, the cross-spectral matrix will have a high number of off-diagonal elements modeling relations between different sources. If the source signals at different positions (indexed by $\ell$ and $\ell'$) are assumed uncorrelated then $S_{\ell \ell'}(\Omt)=0$ for $\ell\neq\ell'$. 
Thus, in this case the previous equation simplifies to
\begin{align}\label{Equ:UniCSpecIndep} 
\nonumber & \gamma_{nn' \ell}(\Om,\Om')=\\
\nonumber &= \frac{1}{v_s^2 2  \pi} \E^{\I \Om v_s^{-1} (x_n - x_\ell)} \E^{-\I \Om' v_s^{-1} (x_{n'} - x_{\ell})}  \sint \q{n\ell}{\Om,\Omt} \qc{n' \ell}{\Om',\Omt}  \E^{-\I\Omt v_s^{-1} (x_n - x_{n'} - x_\ell  + x_{\ell})}  S_{\ell}(\Omt) d\Omt \\
&= \frac{1}{v_s^2 2  \pi} \E^{\I \Om v_s^{-1} (x_n - x_\ell)} \E^{-\I \Om' v_s^{-1} (x_{n'} - x_{\ell})}  \sint \q{n\ell}{\Om,\Omt} \qc{n' \ell}{\Om',\Omt}  \E^{-\I\Omt v_s^{-1} (x_n - x_{n'} )}  S_{\ell}(\Omt) d\Omt, 
\end{align}
where the cross-spectral matrix was replaced by the power spectral density of the process at the $\ell$-th position severely reducing the number of terms to be calculated. In the integrand the dependence on $x_{\ell}$ disappears since the functions $q(\cdot)$ are only dependent on $y$ and $z$-coordinates the longitudinal source position is only included via a phase gradient as long as no cross-terms between different sources are considered (meaning that sources are assumed uncorrelated). Note that a short-hand notation $\gamma_{nn' \ell}$ was introduced to indicate that cross-terms are not considered.
Interestingly, looking at the power and cross-spectral densities observed at the receiver ($\Om=\Om'$) all information about the $x$-coordinate of the source position is lost and only information about the height of the source is retained (under the assumption that all potential sources lie on a plane parallel to the direction of movement):
\begin{align}\label{Equ:UniCrossSpec} 
\nonumber & \gamma_{nn' \ell}(\Om,\Om)=\\
\nonumber &= \frac{1}{v_s^2 2  \pi} \E^{\I \Om v_s^{-1} (x_n - x_\ell)} \E^{-\I \Om v_s^{-1} (x_{n'} - x_{\ell})}  \sint \q{n\ell}{\Om,\Omt} \qc{n' \ell'}{\Om,\Omt}  \E^{-\I\Omt v_s^{-1} (x_n - x_{n'} )}  S_{\ell}(\Omt) d\Omt \\
&= \frac{1}{v_s^2 2  \pi} \E^{\I \Om v_s^{-1} (x_n - x_{n'})}  \sint \q{n\ell}{\Om,\Omt} \qc{n' \ell}{\Om,\Omt}  \E^{-\I\Omt v_s^{-1} (x_n - x_{n'} )}  S_{\ell}(\Omt) d\Omt.
\end{align}
It is important to remember that this last equation is only valid when looking at a single source $\ell$ or at uncorrelated sources. In case of existing correlations between different source positions, the distance between the source positions along the $x$-direction would be contained in the integrand even for $\Om=\Om'$.

\subsection{Locally white process}
One remaining problem with Eq.~(\ref{Equ:UniCSpecIndep}) is the unknown frequency dependency of $S_{\ell}(\Om)$ which needs to be known in order to calculate the integral. 
In this manuscript $S_{\ell}(\Om)$ will be assumed constant in a certain frequency band where the functions $\q{n \ell}{\cdot}$ lead to significant contributions to the integral, thus the term 'locally white'. 
For this it is important to look at the properties of $\q{n\ell}{\cdot}$ for the simplest case, i.e., the moving point source. There
\begin{equation}\label{Equ:25DGreen}
  \q{n\ell}{\Om,\Omt} = \frac{\I}{4} \Ha{0}{ {\sqrt{\Om^2/c^2-(\Om-\Omt)^2/v_s^2}}\sqrt{(y_n-y_\ell)^2 +(z_n-z_\ell)^2}} = \frac{\I}{4} \Ha{0}{k_2 r_{n\ell}}.
\end{equation}
The second term of the argument in Eq.~(\ref{Equ:25DGreen}) is the distance $r_{n \ell}$ between source and microphone positions in the $y$-$z$-plane, which is independent of $\Omt$ and bounded from below by the minimal distance between microphone array and source. The first term can be interpreted as a modified wave number. Also, it is assumed that the speed of sound $c$ is real, i.e., no damping in the medium is given.

The Hankel function has two important properties. First, for purely  imaginary arguments, the Hankel function decays exponentially \cite[Eq.~(9.7.2)]{Abramowitz9}. Second, for small $k_2r_{n \ell}$ the imaginary part of $H_0$ behaves like  $\frac{2}{\pi} \ln(k_2r_{n \ell})$ \cite[Eq.~(9.1.9)]{Abramowitz9} and, thus, tends toward minus infinity for $k_2 \rightarrow 0$. 
For a fixed $\Om$ these ``critical wavenumbers'' where $k_2 = 0$ are given by $\Omt_{\pm} = \Om\!\left(1 \pm v_s c^{-1}\right)$. Since our interest lies in subsonic sources the mach number  $m=v_s c^{-1}$ is real and it is less than one. To be more specific, we usually are interested in cases with speeds of up to a Mach number $m\leq 0.3$, i.e., speeds of up to around 350\,km/h.

\begin{figure}
  \begin{center}
    \includegraphics[trim=0cm 0cm 0cm 0cm, clip=true, width=0.9\textwidth]{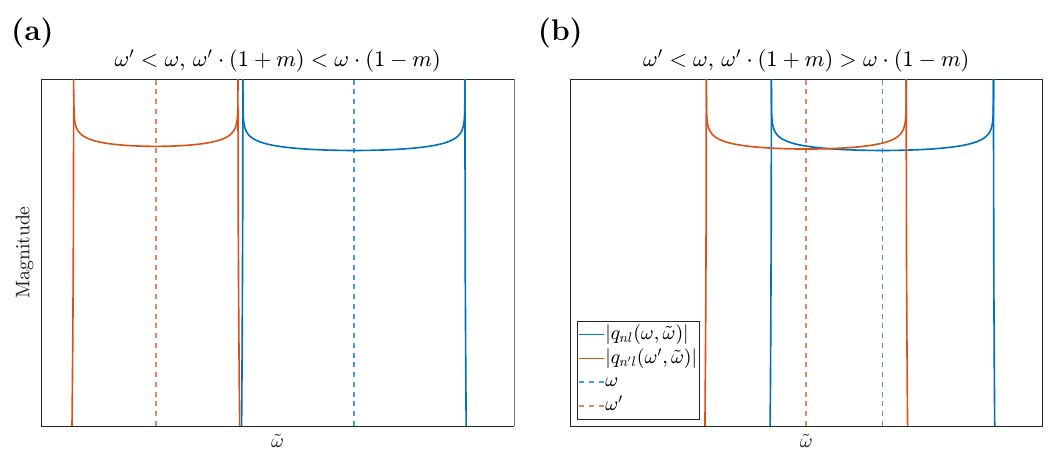}
  \end{center}
  \caption{General properties of the integrand. The graphs illustrate the general behavior of the magnitude of functions $q_{n\ell}$ and $q_{n'\ell}$ for different choices of $\Om$ and $\Omp$. The left panel shows the case where there is no overlap between the non-decaying portions of the functions. The right panel shows a case with some overlap.} \label{Fig:qlns}
\end{figure}

In summary, $\q{n \ell}{\Om,\Omt}$ can be split into three parts. The first part from $(-\infty,\Omt_{-} - \varepsilon)$ where $q_{n \ell}$ decays exponentially, a part between $[\Omt_{-} - \varepsilon, \Omt_{+} + \varepsilon]$ where $q_{n\ell}$ has two (sharp) $\log$-singularities, and a third part between $(\Omt_{+} + \varepsilon,\infty)$ where, again, an exponential decay can be observed (see also Fig.~\ref{Fig:qlns}). As a side note, the singularities only occur in the limit of the undamped case \cite{Duhamel1996}. If a small amount of dampening is introduced such that $\knu = \sqrt{k^2+\I \nu}$, which can be interpreted as a complex speed of sound $\cnu = \omega / \knu$, the singularities disappear (see also Appendix A). Still, the described properties will essentially also hold for the weakly damped case. 

Due to the product in Eq.~(\ref{Equ:UniCSpecIndep}), there are 4 distinct singularities   (two each for $q_{n \ell}$ and $q^*_{n' \ell}$) in Eq.~(\ref{Equ:UniCrossSpec}) unless $\Om=\Om'$. For $\Om$ and $\Om'$ sufficiently apart from each other ($\Om' < \Om, \Om' (1 + m) < \Om (1 - m)$, see Fig.~\ref{Fig:qlns}a) the integral in Eq.~(\ref{Equ:UniCrossSpec}) vanishes numerically because the singularities of $q_{n \ell}$  are numerically ``canceled'' by the exponential decay of $q_{n' \ell}$ and vice versa.
For $\Om$ and $\Om'$ relatively close to each other ($\Om' < \Om, \Om'(1 +
 m) > \Om( 1 - m)$, Fig.~\ref{Fig:qlns}b), $q_{n\ell}$ and $q_{n'\ell}$ partly overlap, where the two ``outer'' singularities are again numerically ``canceled'' by the exponential decay of the respective other function).
Thus, numerically the integrals can be practically limited to the region between the outer two singularities as outside of these both functions $\q{n \ell}{\cdot}$ and $\q{n' \ell}{\cdot}$ exhibit an exponential decay. This is something to be assessed in more detail in the future.

After including the assumption of a stationary and locally white process and omitting any correlations between different source positions, the final equation for calculating the theoretical \Loeve spectrum is:
\begin{align}\label{Equ:UniCSpecIndep_theSecond} 
\gamma_{nn' \ell}(\Om,\Om')&= \frac{S_{\ell}(\Om,\Om')}{v_s^2 (2  \pi)} \E^{\I \Om v_s^{-1} (x_n - x_\ell)} \E^{-\I \Om' v_s^{-1} (x_{n'} - x_{\ell})}  \int\limits_{\Omt_l}^{\Omt_u} \q{n\ell}{\Om,\Omt} \qc{n' \ell}{\Om',\Omt}  \E^{-\I\Omt v_s^{-1} (x_n - x_{n'} )}  d\Omt, 
\end{align}
where $S_{\ell}(\Om,\Om')$ now denotes the power spectral density to be estimated which is assumed to be constant between the integration limits that depend on $\Om$ and $\Om'$ which is indicated by the argument of the source power spectral density. 

 As pointed out before, the source position along $x$ is encoded only via a phase gradient and is not part of the integral. 
Computationally, this leads to a huge gain if a rectangular source grid is used, as the number of vertical grid points determines the number of integrals to be calculated. However, this also leads to the point raised in \cite{Kasess2024} that a particular choice of grid spacing and selecting a regular frequency bin spacing may lead to periodic source distributions along $x$. Thus a suitable choice of $\Om$ and $\Om'$ is again vital and will be addressed in the results section. 

\subsection{Spectral effect of the estimator}
In previous work \cite{Kasess2024} we showed that if one takes  the leakage effect of the windowed DFT into account by modifying the theoretical spectral component accordingly, localization performance can be improved upon. 
In this section it will be investigated, how the averaging properties of the estimator described in Eq.~(\ref{Equ:LoeveEst}) can be included into the theoretical calculation of the \Loeve spectrum.

Employing Eq.~(\ref{Equ:LoeveEst}) but using the theoretical expansion instead of the estimated, the aim is to define the confounding effect of the estimator on the true \Loeve spectrum at receiver pair $n,n'$ caused by source $\ell$: 
\begin{align}\label{Equ:LoeveMTConfound1}
\nonumber \gamma^{MT}_{nn' \ell}[\Om,\Om'] &= \frac{1}{ (2 \pi)^{2}\, K} \EE{\wint\wint 2 \pi \delta(\xi-\xi') d\tilde{Z}_{n \ell}(\Om+\xi) d\tilde{Z}_{n' \ell}^*(\Om'+\xi')}\\
&= \frac{1}{2 \pi \, K}  \wint \EE{d\tilde{Z}_{n \ell}(\Om+\xi) d\tilde{Z}_{n' \ell}^*(\Om'+\xi)}.
\end{align}
Note, that in contrast to the estimates the coefficients in this case are random variables and thus the expected value is used. Using the expansion in Eqs.~(\ref{Equ:Exp1}) and (\ref{Equ:Exp2}) and the orthogonality relation in Eq.~(\ref{Equ:Ortho}) leads to:
\begin{align}\label{Equ:LoeveMTConfound}
\nonumber \gamma^{MT}_{nn' \ell}[\Om,\Om'] &= \frac{1}{ (2 \pi)^{3}\, K} \sum\limits_{k,k'=0}^{K-1}   \wint \wint \wint \nukcc(\eta) \, \nukp(\eta') \EE{dZ_{n \ell}(\Om + \eta) dZ_{n' \ell}^*(\Om' + \eta')} \, \nuk(\xi)  \, \nukpcc(\xi)  d\xi\\
\nonumber &= \frac{1}{ (2 \pi)^{2}\, K }\sum\limits_{k,k'=0}^{K-1} \delta_{kk'}   \wint \wint \nukcc(\eta)  \nukp(\eta') \gamma_{nn' \ell}(\Om + \eta,\Om' + \eta') d\eta d\eta'   \\
\nonumber &= \frac{1}{ (2 \pi)^{2}\, K}\sum\limits_{k=0}^{K-1}  \wint \wint \nukcc(\eta)  \nuk(\eta') \gamma_{nn' \ell}(\Om + \eta,\Om' + \eta') d\eta d\eta' \\
\nonumber &= \frac{1}{ (2 \pi)^{2}\, K}   \wint \wint \sum\limits_{k=0}^{K-1}\left( \nukcc(\eta)  \nuk(\eta')\right) \gamma_{nn' \ell}(\Om + \eta,\Om' + \eta') d\eta d\eta' \\
&= \frac{1}{ (2 \pi)^{2}\, K}   \wint \wint G(\eta,\eta') \, \gamma_{nn' \ell}(\Om + \eta,\Om' + \eta') d\eta d\eta',
\end{align}
where $MT$ indicates that this is the multi-taper 'affected' \Loeve spectrum. In the last two steps the sum was simply pulled into the integral and a weighting function  $G(\eta,\eta')$ was introduced. Essentially, the resulting integral is similar to taking the window properties into account in the single-frequency case. However, first, there are now $K$ windows and, second, the integration is in two dimensions due to the quadratic nature of the estimator. The third difference is more subtle as here the integration is limited by the analysis bandwidth $\mathcal{W}$ whereas in the single-frequency case the integration was, in theory, across the full frequency range. Thus, the leakage outside the bandwidth of interest is ignored. While this cannot be guaranteed, using a restricted number of tapers as described above or adaptive methods to reduce the bias of the multitaper estimates should lead to a good suppression of this leakage effect.
There is an interesting detail about the result in Eq.~(\ref{Equ:LoeveMTConfound}). While the derivation is based on the condition that only the stationary frequency axis is averaged, in the final expression this is not the case. Rather, a rectangular region is averaged across. However, in Sec.~\ref{Sec:Res2} it will be shown that the weighting function $G(\eta,\eta')$ is dominant in a relatively small region around the diagonal.

It is important to note that the confounding effect derived is specific for this particular estimator. Using, e.g., the estimator in \cite{Huang2024}, the confounding effects on the theoretical \Loeve spectrum will be different.

\subsection{Numerical experiment 1}
To evaluate the suitability of the above approach for locating moving stochastic sources, a number of experiments using simulated data is performed.
In a first setting, the moving source is placed at a height of 2\,m and passes through $x_s=0$\,m at time $t=0$. The distance in $y$ is 4\,m and the source is moving at 50 and 100\,\vunit.
For these source parameters, the \Loeve spectrum is derived for a number of source-receiver combinations to show its general properties for $f=(2 \pi)^{-1} \Om= 1000$\,Hz. A range of offsets in $\Delta x_r=x_{n'}-x_n$ and $\Delta z_r=z_{n'}-z_n$ is investigated. The purpose is to illustrate how the relevant frequency range for $\gamma_{nn'\ell}$  w.r.t. $\Om'$ varies for a given setting and a fixed $\Om$ as a function of source speed and horizontal as well as vertical microphone spacing. An approximation to the relevant frequency range in the \Loeve spectrum for realistic array sizes will also be shown.

\begin{figure}[!ht]
\begin{center}
\includegraphics[trim=0cm 0cm 0cm 0cm, clip=true, width=.9\textwidth]{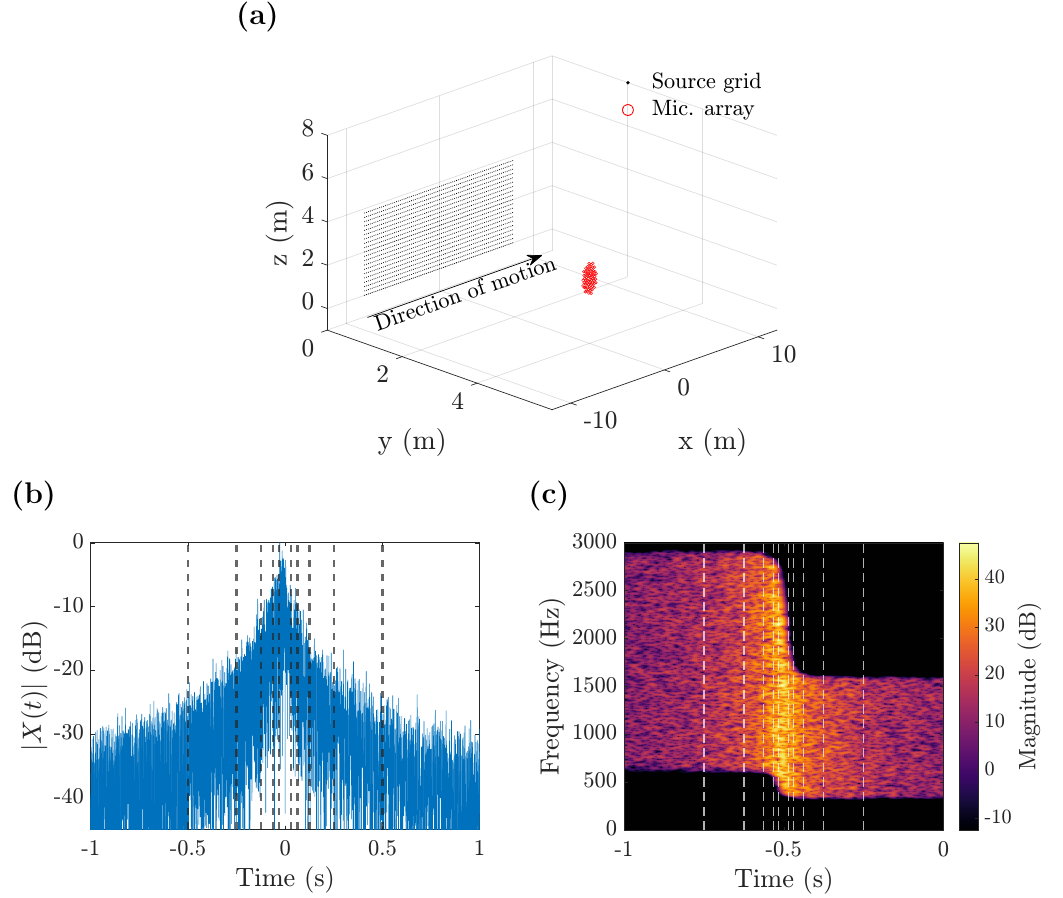}
\caption{{Simulations for numerical experiment 2. Shown is the general scenario (a), signal amplitude (b) and the spectrogram (c) of the first receiver channel as a function of time for $v_s=100$\,\vunit. Vertical lines in panels (b) and (c) mark the different length of tapers.}}
\label{fig:Ex2_sig}
\end{center}
\end{figure}

\subsection{Numerical experiment 2}
Second, an example using a 64-channel microphone array and a moving source will be used to illustrate the suitability of the estimator for the \Loeve spectrum Eq-~(\ref{Equ:LoeveEst}) where $\hat{z}^{(k)}[\Om] := z^{(k)}[\Om]$. The upper panel of Fig.~\ref{fig:Ex2_sig} shows the modeled situation with the array having a distance of 4\,m in the $y$-direction, i.e., to the path of the moving source and the assumed source grid. The receiver geometry is defined by the circular array \textit{tub\_vogel64} with a radius of roughly 1.5\,m from the Acoular package \cite{Sarradj2017}. The speed of the source is again set to 50\,\vunit and 100\,\vunit. The rectangular source grid covers an area of 16$\times$4\,m with a regular grid spacing of 0.2\,m.

The source signals are white noise processes band-limited between 500\,Hz and 2000\,Hz. The effect of the motion of the source is simulated in time domain. This ensures that the generation of the signal and the theoretical calculations are based on two equivalent, but independently implemented approaches. The sampling frequency is set to $f_s=10$\,kHz. Fig.~\ref{fig:Ex2_sig} shows the signal in time domain at a receiver (b) and a spectrogram (c) for a speed of 100\,\vunit.

A number of parameters is investigated to show basic localization properties of the estimator. First, the window length and in turn the number of samples $M$ is varied using values of 62.5\,ms, 125\,ms, 250\,ms, 500\,ms, and 1000\,ms (vertical lines in Fig.~\ref{fig:Ex2_sig}, lower panels). Note that even for 62.5\,ms the time-frequency representation of the signal shows a strong non-stationarity as a considerable frequency transition takes place (innermost pair of white lines). 

The number of tapers $K=2MW$ and hence the bandwidth will also be varied to illustrate the effect of the degree of averaging on the relation between the theoretical and the estimated \Loeve spectrum. Furthermore, the consequences of considering the effect of the tapers on the theoretical \Loeve spectrum given in Eq.~\ref{Equ:LoeveMTConfound} will be investigated. Details will be provided in the results section.

In \cite{Kasess2024} it was shown that the selection of frequencies has a profound effect and that a random choice of frequencies leads to the most reliable results. In the case of the \Loeve spectrum, a number of variants could be considered as two frequencies need to be chosen. The approach chosen here is to keep $\Om$ fixed and randomly vary $\Omp$. The actual range for the choice of $\Omp$ will be explained in the Sec.~\ref{Sec:Res1} of the first experiment. Up to 5 values for $\Omp$ will be chosen for each of the 64$\times$64 receiver pairs.
$S_{\ell}(\Om,\Om')$ in Eq.~(\ref{Equ:UniCSpecIndep_theSecond}) will be set to 1.

The method and the analysis scripts were implemented in MATLAB (R2023a) \cite{MATLAB}. A standard adaptive quadrature was used in this manuscript using the function \texttt{quadgk} from Matlab which uses a Gauss-Kronrod quadrature. Time-frequency plots were generated using the functions \texttt{dgt} and \texttt{plotdgt} of the large time-frequency analysis toolbox (LTFAT, \cite{ltfatnote030}). The DPSS were generated using the MATLAB function  \texttt{dpss}.

\section{Results}\label{sec:results}
\subsection{Numerical experiment 1}
\label{Sec:Res1}
 \begin{figure}[!ht]
\begin{center}
\includegraphics[trim=0cm 0cm 0cm 0cm, clip=true, width=.9\textwidth]{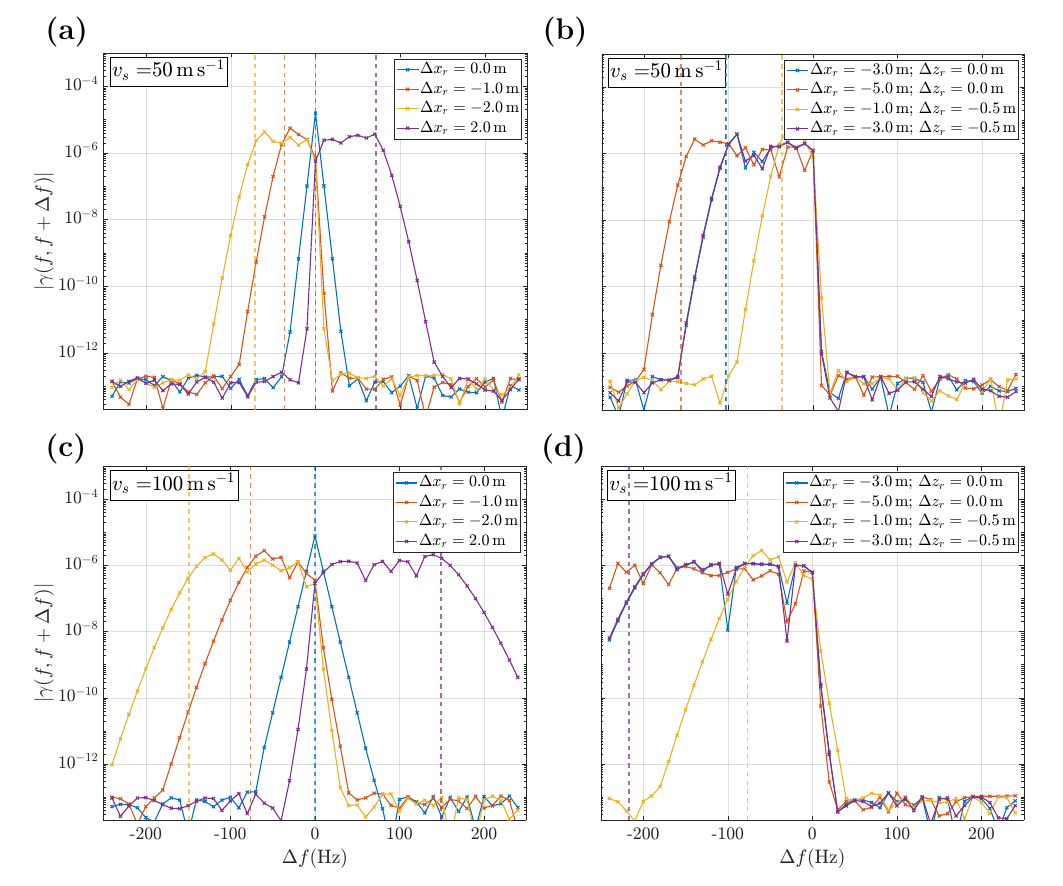}
\caption{{Theoretical \Loeve spectrum as a function of receiver pair distance and frequency shift. Shown is the absolute value of the \Loeve spectrum for a source moving at 50\,\vunit (upper row) and 100\,\vunit (lower row) calculated using Eq.~(\ref{Equ:UniCSpecIndep_theSecond}). Colors code different receiver pairs. Panel (a) and (c) show receiver pairs at the same height $z_r$ whereas in (b) and (d) receivers located at different heights are also shown. The colored vertical dashed lines show the maximum difference in the Doppler-shifted frequency between the two respective receivers that would occur for a single-frequency signal at $1000$\,Hz including the correct sign of the shift. }}
\label{fig:loeve1}
\end{center}
\end{figure}

\begin{figure}[!ht]
\begin{center}
\includegraphics[trim=0cm 0cm 0cm 0cm, clip=true, width=.9\textwidth]{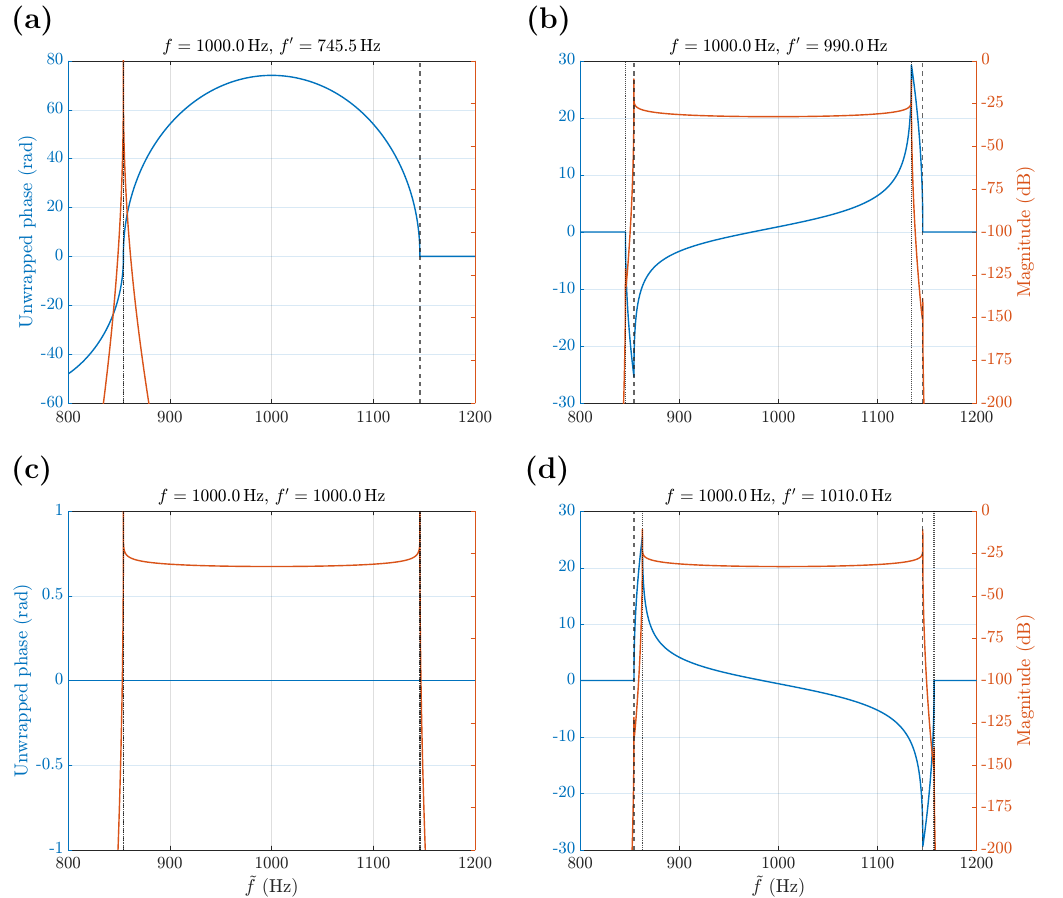}
\caption{Integrand for calculating the \Loeve spectrum as a function of $\tilde{f}$. Shown is the absolute value (orange, right axis) and the phase (blue, left axis) of the integrand of Eq.~(\ref{Equ:UniCSpecIndep_theSecond}) for different combinations of $f$, $f'$, $\Delta x_r=0$, and $v_s=50$\,\vunit. The vertical black dashed lines show the singularities in the portion associated with $f$ and the dotted lines those of the term associated with $f'$. Note that these values can be very close (a) or identical (c). Also note that at this frequency scale, the regions around the singularity cannot be properly resolved. }
\label{fig:integrand1}
\end{center}
\end{figure}

In a first step, the properties of the stochastic process at static observers will be investigated in detail. Using a single source moving from $-\infty$ to $+\infty$ in the free field and a number of different receiver positions, the numerically calculated \Loeve spectrum $\gamma_{nn' \ell}(f,f+\Delta\!f)$ (see Eq.~\ref{Equ:UniCSpecIndep_theSecond}) is displayed in Fig.~\ref{fig:loeve1}. Note that for the remainder of the results section, frequencies in Hertz will be shown rather than angular frequencies, as this is more common and more easily interpretable. Thus, instead of  $\Om$, $\Omp$, and $\Omt$ we will use  $f$, $f'$, and $\tilde{f}$, respectively.
The case of $n=n'$ (i.e., a single observing position) is shown by the blue line in Fig.~\ref{fig:loeve1} for 50\,\vunit (a) and for 100\,\vunit (c). It can be seen that the magnitude of the \Loeve spectrum for $\gamma_{nn \ell}(f,f+\Delta f)$ quickly decays as a function of the frequency difference $\Delta f = f'-f$. Note that the floor at around $10^{-13}$ is due to the tolerance settings and precision of the numerical quadrature. In theory, the spectrum will continue to decay exponentially, but a dynamic range of $10^{-6}$ is sufficient for the purpose.

\begin{figure}[!ht]
\begin{center}
\includegraphics[width=0.8\columnwidth]{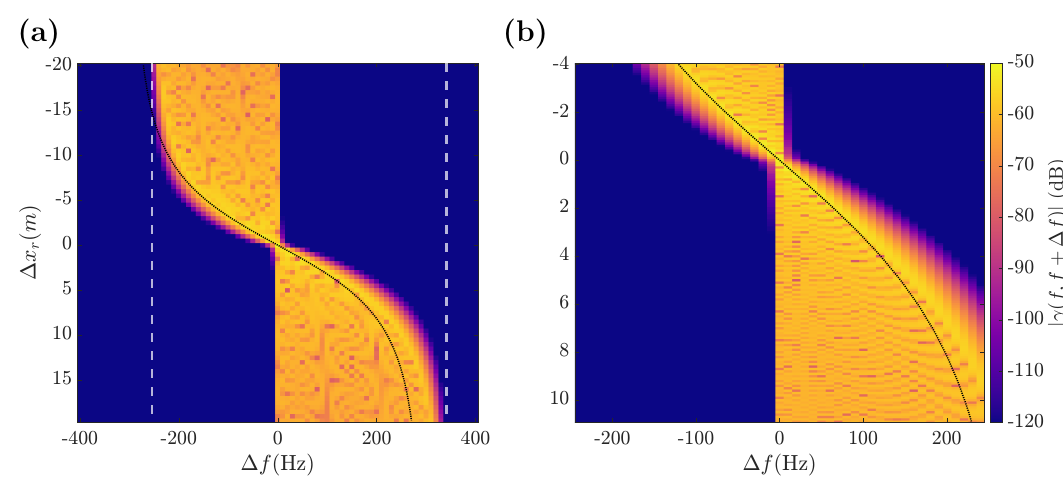}
\caption{Theoretical \Loeve spectrum for large horizontal offsets. (a) Shown is the magnitude in dB of the \Loeve spectrum for receiver pairings with a wide range of $x$-offsets and a fixed frequency $f = 1000$\,Hz. White dashed lines mark the situation were $f' = f (1\pm m)(1\mp m)^{-1}$, i.e., where two singularities coincide. The black dotted line shows the maximum difference in the Doppler-shifted frequency between two horizontally displaced receivers that would occur for a single-frequency signal at $1000$\,Hz. Panel (b) shows more details for a restricted range of $\Delta x_r$. }
\label{fig:loeve2}
\end{center}
\end{figure}

When two different observers are chosen, $\gamma$ extends from the center position either to positive or negative $\Delta f$ depending on whether the $\Delta x_r = x_{n'} - x_n >0$  or $\Delta x_r<0$, respectively (yellow and purple lines in panels (a) and (c)). This reversal of the sign seems intuitively clear, as for $x_{n'} > x_n $ the observer located at $x_{n'}$ always sees a higher Doppler frequency than the observer at $x_n$ and vice versa. From a mathematical point of view, this order-dependent frequency shift originates from a phase gradient reversal in the product of the two functions in Eq.~(\ref{Equ:UniCSpecIndep_theSecond}) as $\q{n\ell}{\cdot}$ and the conjugated $\qc{n'\ell}{\cdot}$ switch order with respect to $\tilde{f}$. Fig.~\ref{fig:integrand1} shows the phase (blue) and the magnitude (orange) of the integrand in Eq.~(\ref{Equ:UniCSpecIndep_theSecond}) for $\Delta x_r=0$ and $v_s=50$\,\vunit. The phase reversal can be clearly seen in the right column where panel (b) illustrates $f'<f$ and panel (d) shows  $f'>f$. Via the inverse Fourier transform this leads to the reversal of the \Loeve spectrum shown in Fig.~\ref{fig:loeve1}.

When comparing different offsets $\Delta x_r$ in Fig.~\ref{fig:loeve1} it is clear that the relevant frequency range in the \Loeve spectrum increases with increasing  $|\Delta x_r|$.
To illustrate the theoretical boundaries, Fig.~\ref{fig:loeve2}(a) shows $|\gamma|$ for a range of horizontal receiver spacings much larger than usually encountered in microphone arrays for close-by sources. The white dashed lines mark the values for $\Delta f$ where $f'$ is selected such that 2 singularities of the Hankel function coincide (see Fig.~\ref{fig:integrand1}(a) were $f'=f (1- m)(1+ m)^{-1}$ at 50\,\vunit). Moving beyond this point leads to a fast decay in the integrands magnitude and, thus, the integral (i.e., the \Loeve spectrum) will also decay quickly, regardless of how large the horizontal receiver spacing is. Importantly, the function is not symmetric w.r.t. to $\Delta f$, although the asymmetry seems not strong for realistic values of $\Delta x_r$.
The black dotted line shows the maximal signed frequency shift  between the two receiver positions that occurs because of the Doppler effect over time while the source is passing. These lines correspond to the dashed vertical lines shown in Fig.~\ref{fig:loeve1} for the specific settings. In contrast to the \Loeve spectrum for a fixed $f$, this quantity is symmetric and thus approximates the boundary observed only for small but realistic  values of $\Delta x_r$  as also shown in a more detail in Fig.~\ref{fig:loeve2}(b). 
The results shown in Figs.~\ref{fig:loeve1} and \ref{fig:loeve2} indicate that for typical microphone arrays this easy-to-calculate quantity provides a suitable selection of frequencies for which the \Loeve spectrum is estimated and will be used in Sec.~\ref{Sec:Res2}.

Finally, while large distances between the receivers along $x$ lead to considerable spectral shifts, the right panels in Fig.~\ref{fig:loeve1} shows that for displacements $\Delta z_r\neq 0$ the effect on $\gamma$ is quite small (c.f. the blue and the purple line which have the same $x$-shift).

\subsection{Numerical experiment 2}
\label{Sec:Res2}
The results in Sec.~\ref{Sec:Res1} show that the range of frequencies for which a contribution is to be expected varies depending on the distance between the receiver pair along the $x$-direction. Thus, for the second setting the values for $f'$ were chosen randomly from the range given by $f$ and the maximal frequency shift that can be seen  between a receiver pair (see the vertical dotted lines Fig.~\ref{fig:loeve1}). This interval was extended by a fixed amount of 20\,Hz for 50\,\vunit (40\,Hz for 100\,\vunit). For $n=n'$, the range was restricted to $\pm$20\,Hz for 50\,\vunit ($\pm$40\,Hz for 100\,\vunit). Frequencies for the \Loeve spectrum were chosen to lie on the DFT frequency grid, i.e., integer multiples of $T^{-1}$. 
Fig.~\ref{fig:Ex2_sig}(b) shows the 'measured' signal at a one receiver position. Vertical dashed lines mark the different values for taper length $T$. It can be seen that the short windows ($T\leq 125$\,ms) do only capture the transient portion of the observed process whereas the longer windows  $T\geq 250$\,ms also capture a, albeit small, portion of the signal caused by the approaching and receding source. 

\subsubsection{Source localization}
\begin{figure}[!ht]
\begin{center}
\includegraphics[trim=0cm 0cm 0cm 0cm, clip=true, width=0.99\textwidth]{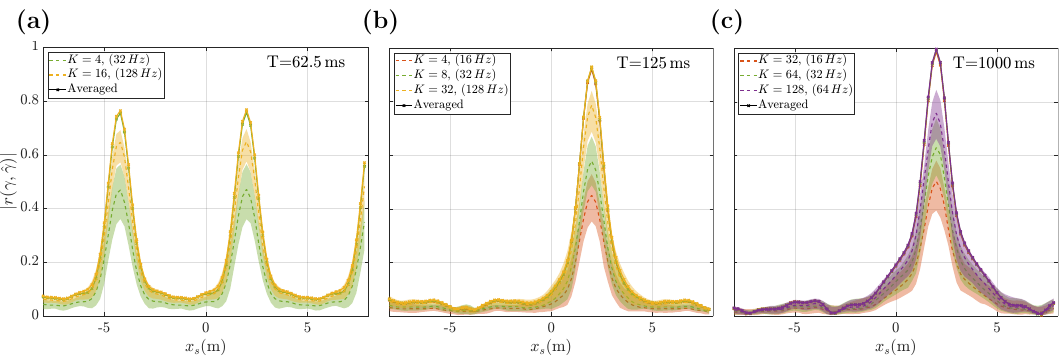}
\caption{{Correlation between theoretical and estimated \Loeve spectrum  along $x$ at the true $z$-position. Panels show the magnitude of the correlation for different values of $T$ as indicated in the upper left corner. Colors denote different number of tapers. Parenthesis in the legend show the bandwidth $W_f$ in Hz. Shaded areas show the 0.05 and 0.95 percentile of the correlations for single realizations, thick dashed lines show the median. Thick solid lines and symbols show the correlation w.r.t. estimates averaged across 100 realizations.  }}\label{fig:Ex2_1_x}
\end{center}
\end{figure}

To evaluate the capability of localizing a moving source with the new approach, Fig.~\ref{fig:Ex2_1_x} shows the absolute value of the correlation $|r(\gamma,\hat{\gamma})|$ between the theoretical and estimated \Loeve spectrum across all $x$-positions of the source grid when the true $z$-position of the source is assumed. Results are shown for a speed $v_s=100$\,\vunit. The shaded areas illustrates the range of variation of the correlation across 100 realizations of the source process.  Results are shown for different taper length $T$ (panels) and different number of tapers $K=2MW$ resulting in different bandwidths (color). For these results, a single value for  $f'$ is used per receiver pair, thus a total of 4096 (64$\times$64) observations were obtained. Note that in contrast to a standard cross-spectral matrix, the two pairings $n,n'$ and $n',n$ are not just complex-conjugated values, as illustrated by the asymmetry in Fig.~\ref{fig:loeve2}. 

In the theoretical result the $x$-direction is only encoded as a phase gradient, and thus the shortest taper $T=62.5$\,ms leads, for a speed of 100\,\vunit, to a period of $T v_s=6.25$\,m in the correlation which is visible due to the horizontal extension of the source grid. For longer tapers the period extends beyond the range of sources and thus the effect of the period is not visible anymore and can be ignored. 
Overall, a main factor determining the strength of the correlation for single pass-bys is the bandwidth.  When comparing across different values for larger values of $T$ (panels (b) and (c) in Fig.~\ref{fig:Ex2_1_x}) similar bandwidths lead to similar ranges of correlation coefficients.
The range in the single pass-by results is a consequence of the variance in the \Loeve estimate across realizations. When averaging the \textit{estimated spectra} across the 100 realizations, the high correlation values w.r.t. the the theoretical value (solid line and symbols) indicate that the estimate converges towards the theoretical calculations. Furthermore, the correlations seem less dependent on the number of tapers when averaging. However, the length of the taper does play a crucial role regarding the strength of the correlation at the true position. Most likely, the reason is that the observation period is too short to cover all the essential properties of the process. It is important at this point to restate that the \Loeve spectrum is a descriptor of the full process, i.e., over the time ranging from $-\infty$ to $+\infty$. Thus, an overly short observation will lead to omissions which in turn explains the lower correlation.
For the lower speed of $v_s=50$\,\vunit all results are very similar an thus for the remainder of the results section only the results for $v_s=100$\,\vunit will be shown. 
Importantly, the spatial periodicity for $v_s=50$\,\vunit is half the length for the same value of $T$ as compared to $v_s=100$\,\vunit. For example, for the shortest taper this leads to a period of $3.125$\,m and thus short tapers pose an even bigger problem for lower speeds.

\begin{figure}[!ht]
\begin{center}
\includegraphics[trim=0cm 0cm 0cm 0cm, clip=true, width=0.99\textwidth]{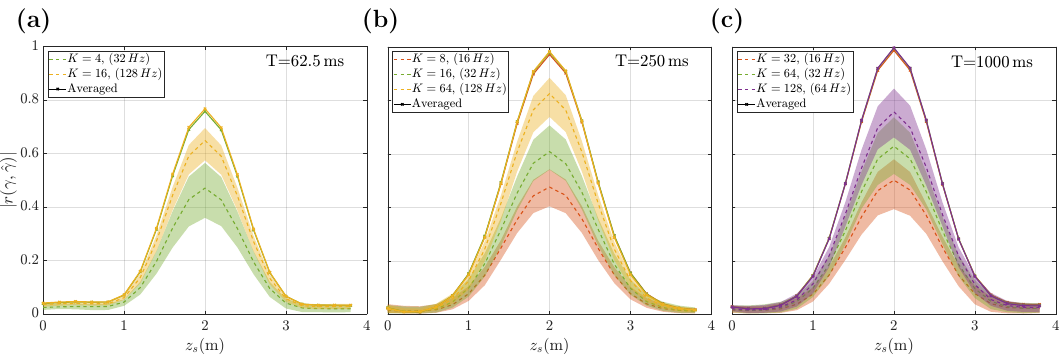}
\caption{{Correlation between theoretical and estimated \Loeve spectrum along $z$ at the true $x$-position. Panels show the magnitude of the correlation for different values of $T$ as indicated in the upper left corner. Colors denote different number of tapers. Parenthesis in the legend show the bandwidth $W_f$ in Hz. Shaded areas show the 0.05 and 0.95 percentile of the correlations for single realizations, thick dashed lines show the median. Thick solid lines and symbols show the correlation w.r.t. estimates averaged across 100 realizations.  }}
\label{fig:Ex2_1_z}
\end{center}
\end{figure}

\begin{figure}[!ht]
\begin{center}
\includegraphics[trim=0cm 0cm 0cm 0cm, clip=true, width=0.99\textwidth]{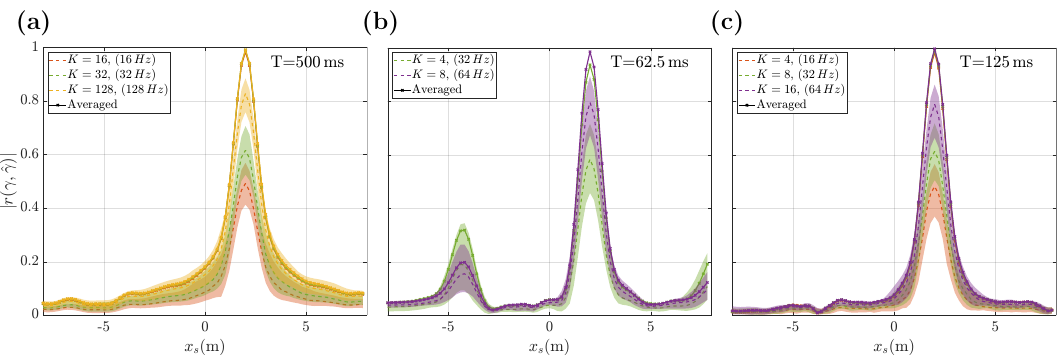}
\caption{{Effect of the number of frequencies per receiver pair and effect of the tapers. Panel (a) shows the correlation when using five values of $\Omt$ per receiver pair. Panels (b) and (c) show the correlation when including the effect of the estimator into the theoretical calculation for different taper lengths. Colors denote different number of tapers. Parenthesis in the legend show the bandwidth $W_f$ in Hz. Shaded areas show the 0.05 and 0.95 percentile of the correlations for single realizations, thick dashed lines show the median. Thick solid lines and symbols show the correlation w.r.t. estimates averaged across 100 realizations.  }}
\label{fig:Ex2_2_x}
\end{center}
\end{figure}

The results for the $z$-direction are very similar (Fig.~\ref{fig:Ex2_1_z}). As the vertical direction is not only encoded in the \Loeve spectrum via a phase gradient, no periodicity is to be expected, and thus the smaller size of the source grid in vertical direction is sufficient to illustrate the basic localization properties. When comparing panels (b) and (c) it can again be seen that similar bandwidths $W_f$ lead to similar correlation coefficients, at least for windows with $T\geq 250$\,ms. Concerning the number of frequencies per receiver pair, Fig.~\ref{fig:Ex2_2_x}(a) shows that the effect of using multiple values for $f'$ does not strongly affect the correlation between theoretical results and the estimates.

\begin{figure}[!ht]
\begin{center}
\includegraphics[trim=0cm 0cm 0cm 0cm, clip=true, width=0.99\textwidth]{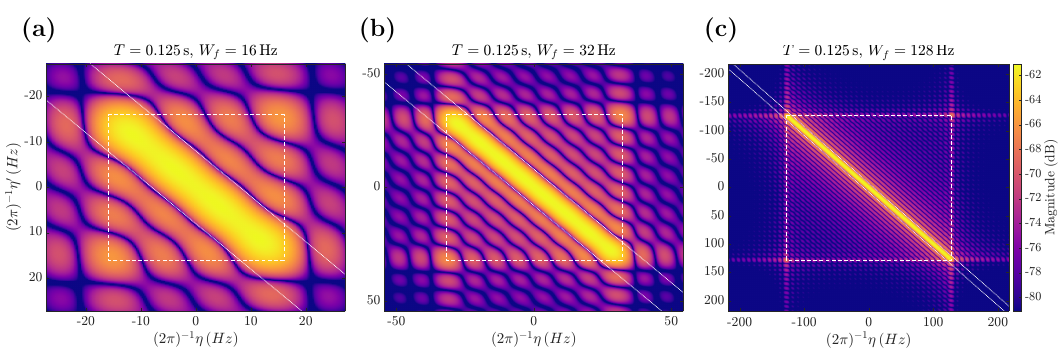}
\caption{{Spectral confounding effect due to the Slepian wave functions. Shown is the magnitude in dB of the weighting function $G(\eta,\eta')$ for the estimator as given in Eq.~(\ref{Equ:LoeveMTConfound}) for $T=125$\,ms and different bandwidths $W_f$ (left to right). The white dashed box illustrates the region of integration as given by the respective value for $W_f$. The white dotted line show delineate the case of $(2 \pi)^{-1} (\eta - \eta')\! =\! \pm T^{-1} = \pm 8$\,Hz.  }}
\label{fig:Confound}
\end{center}
\end{figure}

\begin{figure}[!ht]
\begin{center}
\includegraphics[trim=0cm 0cm 0cm 0cm, clip=true, width=0.99\textwidth]{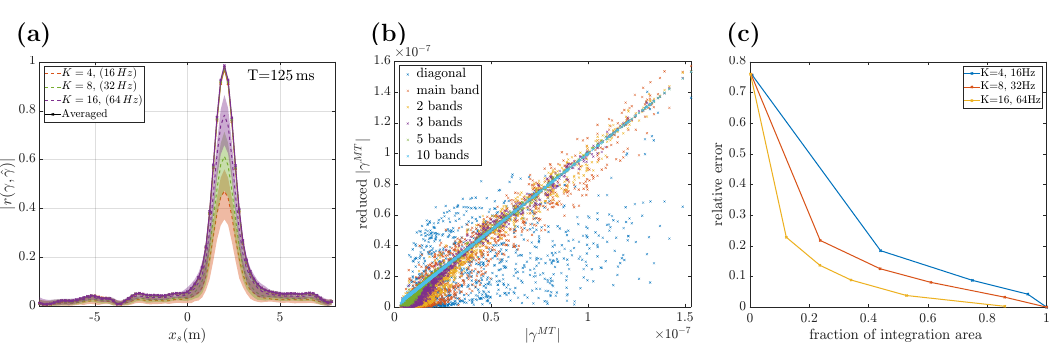}
\caption{{Effect of reducing the integration region. (a) Correlation between theory and estimate when the 2D-integral for the confounding effect is restricted to the main diagonal band. (b) Scatter plot showing the values for the \Loeve spectrum for the reduced regions of integration (y-axis) and using the full square region (x-axis). Colors encode the area of integration. (c) The relative error vs. the relative reduction of the region of integration for different values of $W_f$ (colored dots and lines).}}
\label{fig:Ex2_3_x}
\end{center}
\end{figure}

In the next step, the theoretical \Loeve spectrum confounded by the Slepian wave functions was calculated based on Eq.~(\ref{Equ:LoeveMTConfound}). A rectangular rule and a regular frequency grid was used for $\eta$ and $\eta'$. To keep the computational effort low the maximum bandwidth $W_f$ was 64\,Hz. The maximum number of Slepian wave functions was thus different for different taper lengths. The frequency spacing was set to $2$\,Hz which enables calculations up to $T=500$\,ms. 
The regular grid allows the direct use of the DFT to calculate the Slepian wave functions. For $T<500$\,ms, zero padding was used to obtain the correct frequency spacing of $2$\,Hz.
Fig.~\ref{fig:Ex2_2_x} shows the results for $T=62.5$\,ms (b) and $T=125$\,ms (c). Clearly, for the short tapers the strength of the correlation increases (cf. panels (a) and (b) in Fig.~\ref{fig:Ex2_1_x}). In $x$-direction the periodicity of the correlation for $T=62.5$\,ms is strongly attenuated which is a similar effect as with the method presented in \cite{Kasess2024}. The reason for this effect is that $x_{\ell}$ which occurred only in a phase gradient in the theoretical \Loeve spectrum is now encoded in a different manner due to the additional double-integral. This leads to a disambiguation along the $x$-direction. Overall, the effect on the correlation seems negligible for longer windows as the correlation is barely affected for $T\geq250$\,ms (not shown).
As briefly hinted before, the weighting function $G(\eta,\eta')$ given by the sum of the products of the Slepian wave functions (see Eq.~(\ref{Equ:LoeveMTConfound}) for the expression) has a very specific structure. Fig.~\ref{fig:Confound} illustrates how the confounding effect is concentrated around the diagonal, i.e. $\eta = \eta'$. The width of the main band is given by $\pm T^{-1}$ and is essentially independent of the estimators bandwidth $W_f$, although for small values of $K$ the averaging is not as concentrated. The figures also clearly show the concentration of the averaging process within the square region given by the bandwidth $W_f$ (white dashed box). There is some leakage out of this quadratic region which can be controlled to by reducing the number of tapers from $K=2MW$ to $K=2MW-2$ or even $2MW-4$ tapers since the tapers close to the typical cut-off value already show increased leakage. However, this comes at the cost of decreased averaging and there may be an increased effect of off-diagonal regions within the bandwidth square. 
Overall, while the initial premise of the estimator, that the averaging takes place along the stationary frequency is not strictly fulfilled, the results indicate that the weighting is concentrated around the diagonal. To quantify the effect of the diagonal dominance, the integration was performed using only a specified region around the diagonal. The left panel of Fig.~\ref{fig:Ex2_3_x} shows that when integrating only across the main band of the weighting square, theoretical results already agree, on average, quite well with the estimation (c.f. Fig.~\ref{fig:Ex2_1_x}(b)) although there is still some deviation from the unrestricted integration (c.f. panel (c) Fig.~\ref{fig:Ex2_2_x}). Fig.~\ref{fig:Ex2_3_x}(b) illustrates for $K=16$ how the reduced integration converges towards the unrestricted case when increasing the number of diagonal 'bands' to integrate over. Panel (c) shows the relative error w.r.t. to the unrestricted case as a function of the fraction of the total integration area which in turn is proportional to the squared bandwidth $W_f$. 
The results clearly show that using only the diagonal leads to considerable deviations from the full integral. Furthermore, the periodicity will not be affected in this case. However, the computational cost of considering the confounding effect of the tapers may be reduced to some degree depending on the desired precision.  

\subsubsection{Source separation}

\begin{figure}[!ht]
\begin{center}
\includegraphics[trim=0cm 0cm 0cm 0cm, clip=true, width=0.99\textwidth]{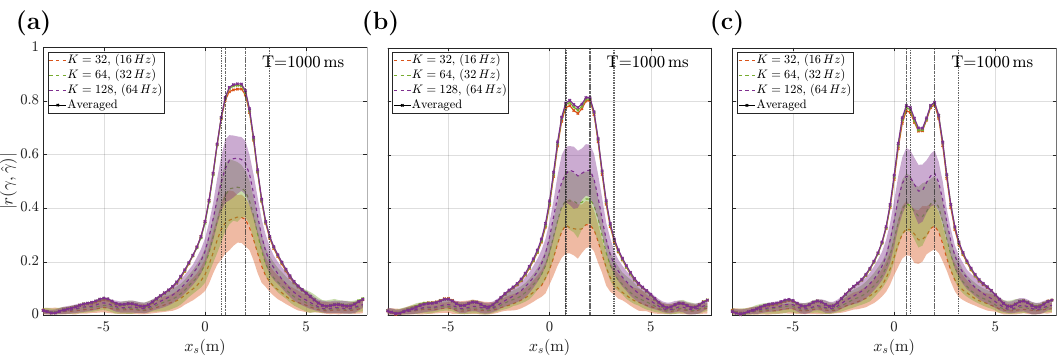}
\caption{{Correlation between theoretical and estimated \Loeve spectrum  along $x$ at the true $z$-position for two sources. Panels show the results for source distances of $\Delta x_s=1$\,m (a), $\Delta x_s=1.2$\,m (b), and $\Delta x_s=1.4$\,m (c) also indicated by the vertical dash-dotted lines. Colors denote different number of tapers. Parenthesis in the legend show the bandwidth $W_f$ in Hz. Shaded areas show the 0.05 and 0.95 percentile of the correlations for single realizations, thick dashed lines show the median. Thick solid lines and symbols show the correlation w.r.t. estimates averaged across 100 realizations. Vertical dotted lines show the Rayleigh distance w.r.t. the source at $x_s=2$\,m. }}
\label{fig:Ex2_1_x_2Src}
\end{center}
\end{figure}
To determine whether the approach is in principle also able to distinguish between adjacent sources, simulated data were generated with two sources at distances of 1\,m up to 2\,m (in steps of 0.2\,m) along $x$ and $z$. The range of distances was chosen as, theoretically, for a wavelength of 0.343\,m at 1000\,Hz, an aperture size (i.e. size of the array) of around 1.5\,m, and a distance of 4\,m between the source and receiver planes the Rayleigh formula leads to a minimal distance between two distinguishable sources of about 1.2\,m  (see, e.g,. \cite{Merino2019}). 
Fig.~\ref{fig:Ex2_1_x_2Src} shows, that the results agree well with the theoretical Rayleigh limit, i.e., on average 2 distinct peaks in the correlation appear although the peaks indicate a potential small reduction in source distance. Including the effect of the tapers on the \Loeve spectrum using Eq.~(\ref{Equ:LoeveMTConfound}) does not yield any changes other than a slightly increased overall correlation for short tapers as observed for the single source case.

\begin{figure}[!ht]
\begin{center}
\includegraphics[trim=0cm 0cm 0cm 0cm, clip=true, width=0.75\textwidth]{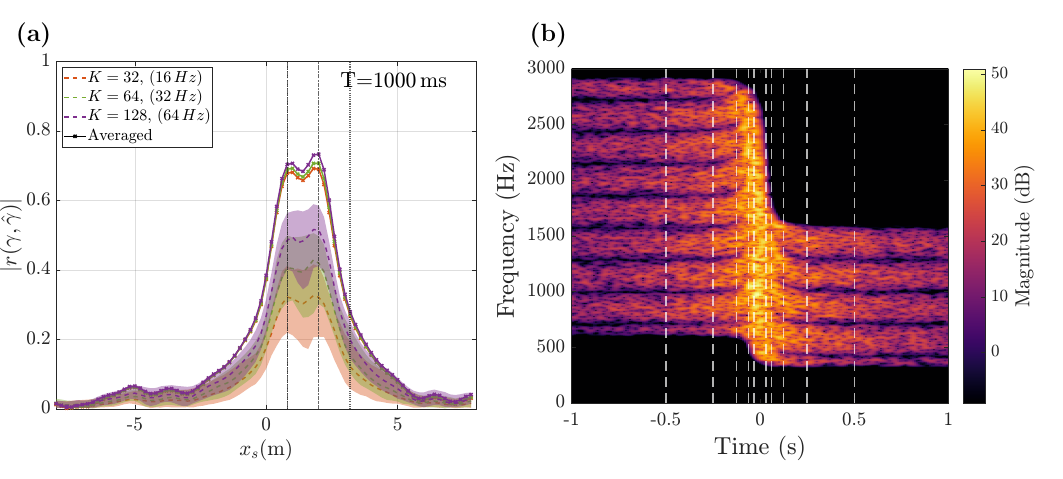}
\caption{{Effect of correlated sources. (a) Correlation between theory and estimate for two perfectly correlated sources at a distance of $\Delta x_s=1.2$\,m. (b) Spectrogram of a receiver channel as a function of time for two correlated sources and $v_s=100$\,\vunit.}}
\label{fig:Ex2_1_x_2SrcCoh}
\end{center}
\end{figure}

When multiple sources are present, the assumption about the uncorrelatedness of different sources may become problematic. Using the same noise for both sources, i.e. introducing a perfect correlation between the two source signals, the same settings were used as in Fig.~\ref{fig:Ex2_1_x_2Src}. Fig.~\ref{fig:Ex2_1_x_2SrcCoh}(a) shows, that the correlation between theory and estimate decreases, although the peaks are still visible on average at the Rayleigh distance (cf. Fig.~\ref{fig:Ex2_1_x_2Src}(b)). Interestingly, due to interference a source distance dependent comb-filter-like effect occurs as can be seen in Fig.~\ref{fig:Ex2_1_x_2SrcCoh}(b), showing that spectral properties can be severely affected for such strong correlations between two sources. This explains the decreased agreement between the theory of uncorrelated sources and the estimate, although the assumption of a full correlation between two separated point sources is clearly a very artificial situation.

\subsubsection{Effect of source spectrum}
The assumption of a constant power spectral density is problematic, as the higher the speed the larger the spectral range that needs to be considered in the integration. In this section, for the high-speed case of $v_s=100$\,\vunit noise filtered using a filter with a spectral shape of a Gaussian curve centered around 1\,kHz and a standard-deviation of 100\,Hz will be considered which strongly affects the spectrum within the integration limits.

\begin{figure}[!ht]
\begin{center}
\includegraphics[trim=0cm 0cm 0cm 0cm, clip=true, width=0.99\textwidth]{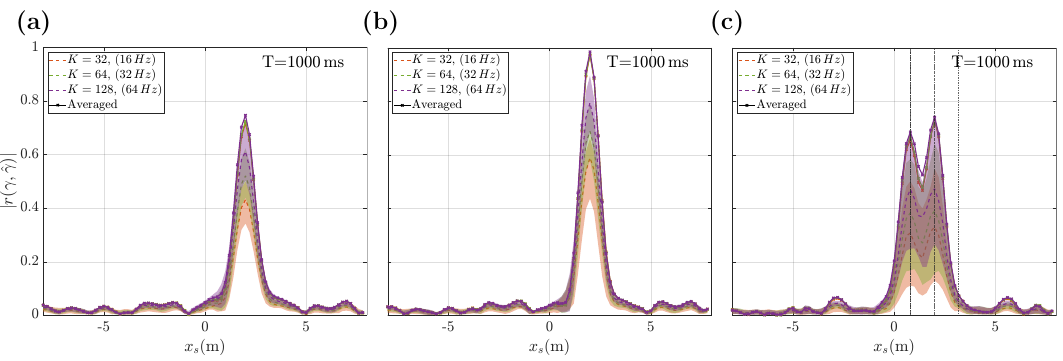}
\caption{{Effect of Gaussian filtered noise. Shown is the correlation for a single source and a taper length of 1000\,ms assuming a constant spectrum (a) and using the true spectral shape for the theoretical \Loeve spectrum (b). (c) The same as (b) for two perfectly uncorrelated sources at a distance of $\Delta x_s=1.2$\,m.  }}
\label{fig:Ex2_1_x_Gauss}
\end{center}
\end{figure}

For this filtered noise (Fig.~\ref{fig:Ex2_1_x_Gauss}) the source is still correctly localized, although the correlation values are lower (a). Taking the known spectral shape into account by using Eq.~(\ref{Equ:UniCSpecIndep}) to calculate the \Loeve spectrum, correlation values increase considerably (b) for a single source. Importantly, in both cases, the localization seems to be sharper, i.e., the base of the peak is slightly narrower which may be a result of less influence of noise from outside the spectral region of interest. 
Looking at Fig.~\ref{fig:Ex2_1_x_Gauss}(c), two sources at the Rayleigh distance lead to two distinct peaks in the correlation. When not taking the spectral shape into account, the curves look similar, although correlation values are consistently lower. 

\section{Discussion}
The main focus of this study was to investigate the transformation of statistical spectral properties of a broad-band stochastic signal emitted by a uniformly moving source using a 2.5D approach and to illustrate whether that information can be used for sound source localization via measurements using a static microphone array. In the proposed method the \Loeve spectrum was used which is a function of two frequencies that captures the non-stationary properties of processes in the spectral domain without explicitly referring to time. The general idea of the presented approach is based on previous work for single-frequency source signals and differs from most other approaches in the field which either aim to transform the non-stationary microphone signals measured in a static frame into the moving reference frame and/or look at a sequence of short snippets which are assumed to be quasi-stationary. 

It was shown that the \Loeve spectrum for uniformly moving sources can be derived in a relatively straight-forward manner within the 2.5D framework. This enables the use of powerful and versatile numerical tools such as the boundary element method to allow the consideration of scattering structures in the vicinity of the microphone array. However, to be numerically feasible, a number of factors was identified to reduce the computational effort, most notably the assumption of a stationary source signal which is related to the main motivation for the proposed approach which is the use along railway tracks. There, a typical measurement set-up is along a straight track with constant track parameters such as track decay rate and rail roughness. Together with the prerequisite of a constant pass-by speed the assumption that the various source processes, which can result, e.g., from the excitation due to the combined wheel and rail-roughness as well as from aerodynamic effects, have constant statistical properties during the pass-by seems well justified. Furthermore, in this setting where the distance between array and sources is relatively small and the speed is high, the main energy of each source will be concentrated in a relatively small slice of time while the source passes the array, even if the entire pass-by itself is comparatively long in the range of a few seconds.

In order to estimate the \Loeve spectrum from microphone signals, an estimator based based on the multitaper approach using the so-called discrete prolate spheroidal sequences was used. The general idea is to estimate spectral properties of a process using a set of orthogonal windows or tapers where each taper covers the temporal range of interest as a whole. The necessary averaging is performed across the different tapers. 

Results based on simulated data for a single source moving at speeds of 50 and 100\,\vunit and emitting bandpass filtered Gaussian white noise showed that the \Loeve spectrum is suitable for localizing a source in the horizontal and vertical direction. The correlation between the estimated and the theoretical \Loeve spectrum showed a clear peak at the true position. When averaging over multiple pass-bys the increasing correlation values indicate that the estimate tends towards the theoretical value, as long as the windows are long enough to cover the essential portions of the pass-by signal. For single pass-bys the multitaper approach displayed a trend towards a better agreement for higher bandwidths $W$ and thus increased averaging due to the higher number $K$ of tapers used. 
It was also shown that for a single stationary source using the \Loeve spectrum leads to an inherent periodicity in the correlation between theory and estimate along the horizontal direction depending on the taper length and the source speed. This is due to the pure phase coding of the $x$-position which may be visible if the field-of-view for the source grid extends beyond the length given by the product of the speed and the taper length. 
Finally, increasing the number of observations per receiver pair did not lead to large changes in the correlation although this may become important if the dimension of the source grid and thus the number of unknown sources is increased. 
For two adjacent sources at and beyond the Rayleigh distance on average two distinct correlation peaks were observed, although the shape of the correlation function indicates heavy blurring. Importantly, looking at the correlation in this way is very similar to conventional beamforming where the projection onto different steering vectors is evaluated. As pointed out in the introduction, the resolution of beamforming can be greatly enhanced using various deconvolution methods as is the case when using suitable regularization methods for inverse approaches. Thus, using appropriate inversion methods for the proposed forward model will be crucial to lead to better resolution for point-like sources.

Since the spectral averaging is inherent to the estimate but not the theoretical \Loeve spectrum, a method was derived to take the effect of the bandwidth of the tapers on the theoretical values into account. Simulation results illustrate that this leads to a better agreement with the estimate, in particular for short tapers. Visible periodicity effects that occur for tapers shorter than the time it takes to traverse the length of the source grid could be partially suppressed. However, this better agreement comes at the cost of a severely increased computational effort which can be partially reduced by restricting the two-dimensional integration to a region centered around the diagonal. In more realistic scenarios where sources are distributed over a larger area such short segments of the recordings are most likely not suitable to estimate the \Loeve spectrum as this would not allow to cover the essential parts of the signal. For longer segments ignoring the taper effect is most likely not overly problematic, at least for source localization. 

In addition to stationarity of the source, another major assumption about the sources was that the source spectrum is essentially flat over a certain frequency range that depends on the speed. Clearly, in realistic scenarios a flat spectrum over a large spectral range cannot be assumed. Using simulated data where this assumption was not fulfilled lead to a decreased agreement in terms of the correlation coefficients although the peak of the correlation was still at the true position and the case of two sources lead to qualitatively comparable results. When the spectral shape was taken into account in the theoretical calculation, the agreement was similar as in the spectrally flat case.

The third simplification was to ignore cross-correlation terms between different sources, i.e. to assume that sources can be treated as uncorrelated. Using simulated data of two fully correlated sources showed, that the presence of correlations can affect the results, at least in terms of the overall agreement with the theory that assumes uncorrelated sources. 
While the scenario used was artificial and an extreme case, it is clear that correlations between close-by sources may occur. For example, the wheel of a railway vehicle radiates with different modes and thus different points on the wheel will be correlated depending on the frequency.

\section{Conclusion}
Summarizing, the results presented indicate the potential for directly using quantities describing properties of non-stationary processes for source localization for uniformly moving sources in a 2.5D framework. The versatility of 2.5D methods makes this a powerful approach for localization.
Focusing on the forward model, a suitable inversion scheme was not covered in this work but will be investigated in the future. In theory, among others, any regularized inversion method may be used. Future work will also be directed towards loosening the restrictions used in the derivation of the theoretical results. 
One important step is the modeling of the spectral shape in functional form to allow concurrent estimation of a larger frequency range. In addition, the feasibility of modeling at least spatially limited correlations between sources will be investigated. 
Another important aspect of future work will be the application of other estimators from the literature which will require to derive the spectral effect on the theoretical \Loeve spectrum.

\section*{Acknowledgements}
This work was supported by the Austrian Science Fund (FWF) via the WEAVE project LION (Localization and Identification of moving noise sources, DOI: 10.55776/PIN9082923).

\section*{Appendix A}
In this section details are provided to illustrate the introduction of a harmonizable stochastic process into the 2.5D framework.
The starting point is Eq.~(\ref{Equ:UniInd2}), which is reproduced here:
\begin{align}\label{Equ:UniInd2A} 
 p_{n\ell}(t) &=  \frac{1}{v_s (2 \pi)^2} \dint  \shat_\ell(\Omt) \q{n\ell}{\sqrt{\Om^2/c^2-(\Om-\Omt)^2/v_s^2}}  \E^{\I (\Om-\Omt) v_s^{-1} \Delta x_{n\ell}} \E^{-\I \Om t} d\Omt  d\Om.
\end{align}
A shorthand notation $\Delta x_{n\ell}$ is used for $x_{r,n} - x_{s,\ell}$. The spectrum $\shat_\ell(\Omt)$ can be represented via a Fourier transform of $s_\ell(t)$:
\begin{align}\label{Equ:UniInd3A} 
 p_{n\ell}(t) &= \frac{1}{v_s (2 \pi)^2} \dint \left(\sint s_\ell(t') \E^{\I \Omt t'} dt' \right) \q{n\ell}{\sqrt{\Om^2/c^2-(\Om-\Omt)^2/v_s^2}}  \E^{\I (\Om-\Omt) v_s^{-1} \Delta x_{n\ell}} \E^{-\I \Om t} d\Omt  d\Om.
\end{align}
Unfortunately, the stochastic process cannot be introduced directly as the Fourier transform can only be defined in a weak sense (see, e.g., \cite[Theorem 3.1]{Radchenko2017}):
\begin{align}\label{Equ:Inversion} 
 \sint \phi(\Omt) dY(\Omt) &= \Plim{\alpha \rightarrow 0^+} \sint \phi(\Omt) \left( \sint \E^{-4\pi^2\alpha |t|^2} \E^{\I \Omt t} X(t) dt\right) d\Omt,
\end{align}
where $\Plim{}$ denotes a limit in probability where the Gaussian window function goes to 1 as $\alpha$ approaches 0. The function $\phi(\Omt)$ needs to have compact support and needs to be infinitely differentiable.
Comparing Eq.~(\ref{Equ:UniInd3A}) and Eq.~(\ref{Equ:Inversion}) and leaving out all terms not containing $\Omt$ and the integral over $\Om$ Eq.~(\ref{Equ:Inversion}) can be rewritten as:
\begin{align}\label{Equ:InversionII} 
 \sint \phi_{n\ell}(\Omt) dY_{\ell}(\Omt) &= \Plim{\alpha \rightarrow 0^+} \sint \q{n\ell}{\sqrt{\Om^2/c^2-(\Om-\Omt)^2/v_s^2}}  \E^{-\I \Omt v_s^{-1} \Delta x_{n\ell}} \left( \sint \E^{-4\pi^2\alpha |t|^2} \E^{\I \Omt t} X(t) dt\right) d\Omt.
\end{align}
The main question is whether $\phi_{n\ell}(\Omt) = \q{n\ell}{\sqrt{\Om^2/c^2-(\Om-\Omt)^2/v_s^2}}  \E^{-\I \Omt v_s^{-1} \Delta x_{n\ell}}$ fulfills the requirements. Remembering that, e.g., for a single moving point source $q_{n\ell}$ is given by $\Ha{0}{r_{n\ell} \sqrt{\Om^2/c^2-(\Om-\Omt)^2/v_s^2}}$ it is clear that for $\Om \in \mathbb{R}$ the function $\phi_{n\ell}(\Omt)$ has neither finite support nor is it infinitely differentiable at the singularities. 
Regarding the support it is important to remember that here only band-limited process are considered. Thus, we can always introduce a weighting function $h(\Omt)$ such that $\phi_{n\ell}(\Omt) \rightarrow h(\Omt) \phi_{n\ell}(\Omt)$ that has the value 1 within the limited band and that goes to zero in a manner that leaves $h(\Omt)$ infinitely differentiable except at the singularities where the argument of the Hankel function $\Ha{0}{.}$ is zero. 
Concerning the differentiability at these points, the Hankel function fulfills the criterion if a small imaginary part is introduce for $k=\omega/c$ such that $\knu = \sqrt{k^2+\I \nu}$ with $\nu > 0$. This can be interpreted as a complex speed of sound $\cnu = \omega / \knu$ leading to a damping if the square root with a positive imaginary value is used such that the Hankel function is evaluated within the first quadrant and the singularity at 0 is omitted. The Fourier transform pair for the Green's function can then be defined in the limit of $\nu \rightarrow 0$ \cite{Duhamel1996}. Thus, strictly speaking Eq.~(\ref{Equ:UniInd3A}) is of the following form when introducing a harmonizable stochastic process:
\begin{align}\label{Equ:UniLimiting} 
\nonumber X_{n\ell}(t) &= \frac{1}{v_s (2 \pi)^2} \Plim{\nu \rightarrow 0^+}\!\! \left[ \sint \E^{-\I \Om t} \E^{\I \Om v_s^{-1} \Delta x_{n\ell}} \, \sint  \Plim{\alpha \rightarrow 0^+}\!\! \sint h(\Omt)\, \q{n\ell}{\sqrt{\Om^2/\cnu^2-(\Om-\Omt)^2/v_s^2}}  \E^{-\I \Omt v_s^{-1} \Delta x_{n\ell}} \right.\\
 \nonumber & \left. \left( \sint \E^{-4\pi^2\alpha |t'|^2} \E^{\I \Omt t'} X_\ell(t) dt\right) d\Omt  d\Om \right] \\
 &= \frac{1}{v_s (2 \pi)^2} \Plim{\nu \rightarrow 0^+}\!\! \left[ \sint \E^{-\I \Om t} \E^{\I \Om v_s^{-1} \Delta x_{n\ell}}  \, \sint  \q{n\ell}{\sqrt{\Om^2/\cnu^2-(\Om-\Omt)^2/v_s^2}}  \E^{-\I \Omt v_s^{-1} \Delta x_{n\ell}} dY_\ell(\Omt) d\Om \right].
\end{align}
where in the last line the weighting function was omitted since $dY_\ell(\Omt)$ is band limited anyway and thus $h(\Omt)$ has, by definition, no effect on the integral within the brackets.

 \bibliographystyle{abbrv}

\end{document}